# MX precipitate behavior in an irradiated advanced Fe-9Cr steel: Self-ion irradiation effects on phase stability

T.M. Kelsy Green[a,1*], Tim Graening[b], Weicheng Zhong[b], Ying Yang[b], and Kevin G. Field[a]

[a]University of Michigan-Ann Arbor
[b]Oak Ridge National Laboratory

[1]Currently at Antares Nuclear Industries

*Corresponding Author: kelsy@antaresindustries.com

## Graphical Abstract

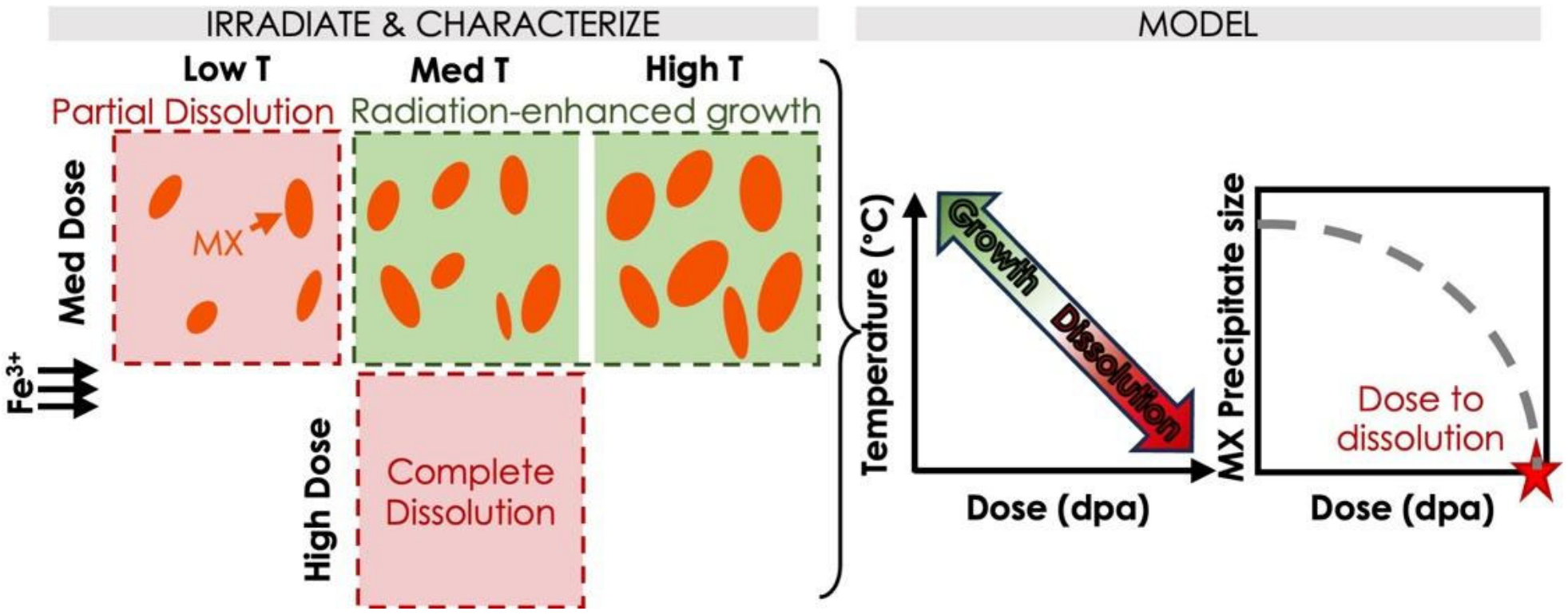

## Abstract

Reduced activation ferritic/martensitic (RAFM) steels are the leading candidate structural materials for first-wall and blanket components in fusion reactors. This work is the first in a series to provide a systematic roadmap of MX precipitate stability in RAFM steels under various ion irradiation conditions. Here, the MX-TiC precipitate behavior in an advanced Fe-9Cr RAFM steel is assessed under self-ion irradiation to damage levels ranging from 1 to 100 displacements per atom (dpa) at temperatures ranging from 300-600°C to isolate the effects of temperature and damage level on precipitate stability. The pre-existing MX-TiC precipitates are shown to exhibit temperature-dominated responses, including coarsening above 400°C at damage levels of 15 dpa, while damage levels studied at 50 dpa and higher showed dissolution across all temperature ranges studied. The effects of ballistic dissolution and diffusion on precipitate behavior are outlined as a function of precipitate characteristics (number density, size, and volume fraction) and irradiation parameters with the use of the recoil resolution model of precipitate stability. This work provides critical insights into MX-TiC stability to high doses in-order to further optimize advanced steels with improved radiation resistance.

### 1.1 Introduction

Designing and manufacturing structural materials that can operate optimally in the harsh environment of a fusion reactor core for long life spans remains a hurdle for fusion energy deployment. Reduced activation ferritic/martensitic (RAFM) steels are the favored candidate structural materials for first-wall and blanket components in fusion reactors [1, 2] due to their inherent microstructural complexity, leading to the resistance of radiation-induced degradation [3]. RAFM steels are body-centered cubic (BCC) Fe-based alloys with 7-15 weight % Cr and 1-3 weight % minor alloying solutes [4]. The inherent complexity of RAFM steels is derived from their grain structure, containing prior austenite grains, subgrains, and martensitic laths; their secondary phases, such as $M_{23}C_6$ and MX precipitates; and their high, pre-existing dislocation densities (~$10^{14}$ $m^{-2}$) [5].

Of particular strategic importance for improving the high temperature creep strength [6] and increasing the sink strength of RAFM steels [7, 8] is the inclusion and stability of MX (M=metal, X=C and/or N) precipitates, found on and within grain boundaries. Extensive studies in Ti-modified austenitic steels have observed that the stability of Ti-rich MX precipitates is controlled by an interplay of concentration gradients, recoil resolution (also known as ballistic dissolution), temperature, and processing of steel prior to irradiation [9-19]. However, a notable gap exists in the systematic documentation of the stability of MX precipitates in traditional ferritic/martensitic (FM) and RAFM steels, especially under conditions relevant to fusion environments [20-38]. As the stability of these precipitates is crucial to maintain creep and sink strength during the entire operation of a fusion reactor (>100 dpa) [39-42], a comprehensive analysis of the behavior of MX precipitates is imperative. Furthermore, a nuanced understanding of MX precipitate behavior will provide insights into the intricate relationship between precipitation and swelling in future work.

This work is the first in a series of three papers to provide a systematic evaluation of MX precipitate behavior in RAFM steels under various ion irradiation conditions, culminating in understanding MX precipitate stability and swelling in dual ion irradiations. This work will look at the precipitate stability under single beam self-ion irradiation to isolate the effects of temperature and damage level on MX-TiC precipitate stability. A combination of ion irradiation experiments, with careful post-irradiation characterization techniques coupled with precipitate stability models, were used to achieve this objective.

## 1.2 Methods

### 1.2.1 Material

The alloy used in this work is part of the new family of Castable Nanostructured Alloys (CNAs), developed as part of the U.S.-led effort to further optimize RAFM steels for fusion applications by increasing MX precipitate number densities (>$10^{21-22}$ $m^{-3}$) with compositional modifications and thermomechanical treatments [4, 43-46]. This innovative approach has led to enhanced mechanical properties and radiation tolerance compared to preceding generations of FM and RAFM steels [1, 2, 4, 19, 43, 47-49]. The CNA steel used in this work is referred to as CNA9.

It is a simple engineering analogue for more complex CNAs and was specifically developed to study the effects of carbon content on the microstructural evolution in CNAs. Through carbon content reduction and careful heat treatment design, CNA9 does not contain $M_{23}C_6$ precipitates. Hence, this low-carbon CNA steel was opted for this investigation over alternative CNA variants to isolate the irradiation effects on the MX precipitates and to avoid any influence from the $M_{23}C_6$ precipitates. CNA9 contains two TiC precipitate populations: small, elliptical MX-TiC precipitates with a diameter range of 3-18 nm (mean equivalent diameter = 7.9±0.3 nm) and large, spherical TiC precipitates with a diameter range of 50-100 nm. The small, elliptical MX-TiC precipitates are of interest in this work as they are hypothesized to be primary contributors to increased creep strength and radiation tolerance in the CNA variants.

The composition of CNA9 in weight percent is reported in Table 1. The bulk chemistry evaluation was performed by Dirats Laboratories using quantitative analysis by inductively coupled plasma spectroscopy optical emission spectroscopy (ICP OES) for metallic elements, combustion for C and S, and inert gas fusion (IGF) for O and N. According to Dirats Laboratories, the uncertainty of each measurement was determined to be either 1% of the absolute composition value or two in the last reported digit, whichever is least. This method was used to determine the error of the analysis in Table 1. The alloy and compositional information were provided by ORNL.

CNA9 was normalized at 1050°C for 1 h in Ar atmosphere then hot-rolled at 1050°C to 0.3 inch-thick with ~21% reduction per pass. It was briefly reheated at 1050°C if needed followed by water quenching. The laboratory-scale ingot of CNA9 was then cut into two halves, top and bottom. The bottom half, which was used in this work, was tempered at 750°C for 30 min, air cooled, and de-scaled. A 1 cm piece from the top half was used for composition analysis [40]. Samples in the form of 1.5 × 1.5 × 10 $mm^3$ bars were cut from the bottom half using electrical discharge machining (EDM) and polished with standard procedures, down to a final surface finish obtained using electropolishing. The electropolishing solution was 10% perchloric acid solution, 90% methanol solution, cooled to -45°C using a methanol bath with dry ice.

Table 1. Chemical compositions (wt%) of CNA9 provided by Dirats Laboratories.

| Element | CNA9 (wt.%) |
|---|---|
| Fe | 89.27±0.02 |
| Cr | 8.688±0.08688 |
| W | 1.026±0.0126 |
| Mn | 0.516±0.00516 |
| Si | 0.141±0.00141 |
| Ta | 0.090±0.0009 |
| Ti | 0.141±0.00141 |
| V | 0.049±0.00049 |
| C | 0.049±0.00049 |
| Al | <0.002±0.0002 |
| B | <0.0005±0.00005 |
| Co | <0.005±0.0005 |
| Cu | <0.002±0.0002 |
| Mo | 0.004±0.0004 |
| Nb | <0.002±0.0004 |
| Ni | <0.007±0.0007 |
| P | 0.004±0.0004 |
| Zr | <0.002±0.0002 |
| S | 0.002±0.0002 |
| O | 0.0012±0.00012 |
| N | 0.0013±0.00013 |

### 1.2.2 Ion irradiation Experiments

The ex-situ ion irradiation experiments were conducted with 9 MeV $Fe^{3+}$ ions at the University of Michigan-Ann Arbor’s Michigan Ion Beam Laboratory (MIBL). The experiments tested the single and combined effects of temperature and damage level and will be discussed in three series: (1) a damage level series at constant temperature (1-100 dpa, 500°C), (2) a temperature series at constant intermediate damage level (15 dpa, 300-600°C), and (3) a temperature series at constant high damage level (50 dpa, 300 and 500°C) (Table 2, Figure 1). The design of these irradiation series aimed to evaluate precipitate behavior under conditions relevant to fusion operation [37, 44], strategically spaced at damage level and temperature intervals conducive to conducting fundamental studies on precipitate behavior. The target matrix damage rate was $7\times10^{-4}$ dpa/s, which is a commonly used dose rate for self-ion irradiation of FM and RAFM steels [45, 50]. Target matrix damage levels ranging from 1 to 100 dpa were used to gain a granular understanding of precipitate evolution. The target temperatures of 300, 400, 500, and 600ºC were used to encompass fusion-relevant temperatures and allow for a detailed understanding of operational mechanisms of precipitate stability as a function of thermal effects at the intermediate dose of 15 dpa. The elevated dose condition of 50 dpa at 300 and 500°C was used to ensure that temperature effects were not significant factors at elevated damage levels based on the initial findings found in the extended dose series component of this work.

Table 2 Ex-situ self-ion irradiation parameters. All irradiations were completed with 9 MeV $Fe^{3+}$. The cells are filled as follows: target parameter/achieved parameter experimentally. For simplicity, the target experimental parameters for each irradiation will be shown in the rest of this work. $T_{irr}$ = temperature of irradiation.

| $T_{irr}$ (°C) | Total dpa | Dose rate (dpa/s) |
|---|---|---|
| 300/298.2 | 15/15.1 | $7\times10^{-4}/7\times10^{-4}$ |
| 300/294 | 50/50.4 | $7\times10^{-4}/7\times10^{-4}$ |
| 400/399 | 15/15 | $7\times10^{-4}/7.2\times10^{-4}$ |
| 500/503 | 1/1.1 | $7\times10^{-4}/7.4\times10^{-4}$ |
| 500/502 | 5/5.3 | $7\times10^{-4}/6.8\times10^{-4}$ |
| 500/500.6 | 15/15 | $7\times10^{-4}/7.1\times10^{-4}$ |
| 500/499.8 | 50/50 | $7\times10^{-4}/7.3\times10^{-4}$ |
| 500/499.6 | 100/100 | $7\times10^{-4}/7.4\times10^{-4}$ |
| 600/601 | 15/15.4 | $7\times10^{-4}/6.9\times10^{-4}$ |

The "quick" Kinchin-Pease (KP) mode in the Stopping and Range of Ions in Matter (SRIM 2013) [51] was used to calculate the depth-dependent damage in CNA9 from the 9 MeV $Fe^{3+}$ ions [52, 53]. Figure 2 shows example damage profiles calculated for the 15 and 50 dpa irradiations. Temperature, pressure, beam current, and beam profiles were monitored during the irradiation experiments. Temperature was controlled to within ±10ºC of the target value during experiments with the use of an infrared thermal pyrometer and thermocouple readings. The beam current was controlled to within ±10% of the desired current resulting in a damage rate of $7\times10^{-4}$ dpa/s for all irradiations. It is important to point out that the damage level induced in the CNA9 matrix will be different than that in the MX-TiC precipitates due to the differences in composition and structure between the matrix and the precipitates. To accurately assess the evolution of precipitates with irradiation, the damage level induced in the MX-TiC precipitates was also calculated. The details of how this was calculated with the use of SRIM can be found in Supplemental A. It was determined that the damage level in the MX-TiC precipitates was ~0.467× less than the level in the CNA9 matrix in the nominal damage region shown in Figure 2. The result is the following dose relationships between the matrix and the precipitates: $0.5_{TiC}/1_{matrix}$ dpa, $2.3_{TiC}/5_{matrix}$ dpa, $7_{TiC}/15_{matrix}$ dpa, $23_{TiC}/50_{matrix}$ dpa and $46_{TiC}/100_{matrix}$ dpa.

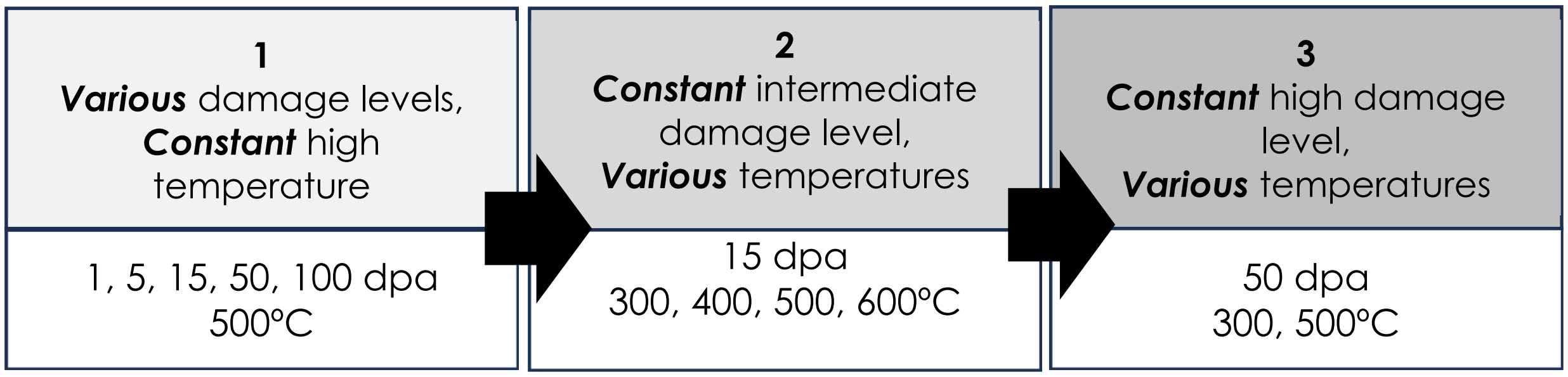

Figure SEQ Figure * ARABIC 1 Graphical representation of the irradiation experiments conducted to understand the effects of irradiation on MX phase stability. All irradiations were conducted with a damage rate of $7\times10^{-4}$ dpa/s.

### 1.2.3 Sample preparation and characterization

Conventional and scanning transmission electron microscopy (S/TEM) was used to analyze the MX-TiC precipitates. A standard TEM sample preparation methodology was used to create lift-outs for S/TEM analysis. Energy-filtered TEM (EFTEM)-based procedures [54] on a Thermo Fisher Tecnai G2 F30 TWIN Electron Microscope (TF30) equipped with a Gatan® Imaging Filter (GIF) were conducted on lift-outs to obtain the lift-out thicknesses. MX-TiC precipitates were identified through elemental mapping using STEM equipped with Energy Dispersive X-ray Spectroscopy (EDS) capability on the Thermo Fisher Talos F200X G2 S/TEM using the high-visibility low-background double-tilt ($\alpha = \pm 35°$, $\beta = \pm 30°$) holder for optimized EDS acquisition. Data acquisition was completed using Velox software. All microscopes used were part of the Michigan Center for Materials Characterization ($(MC)^2$) facility.

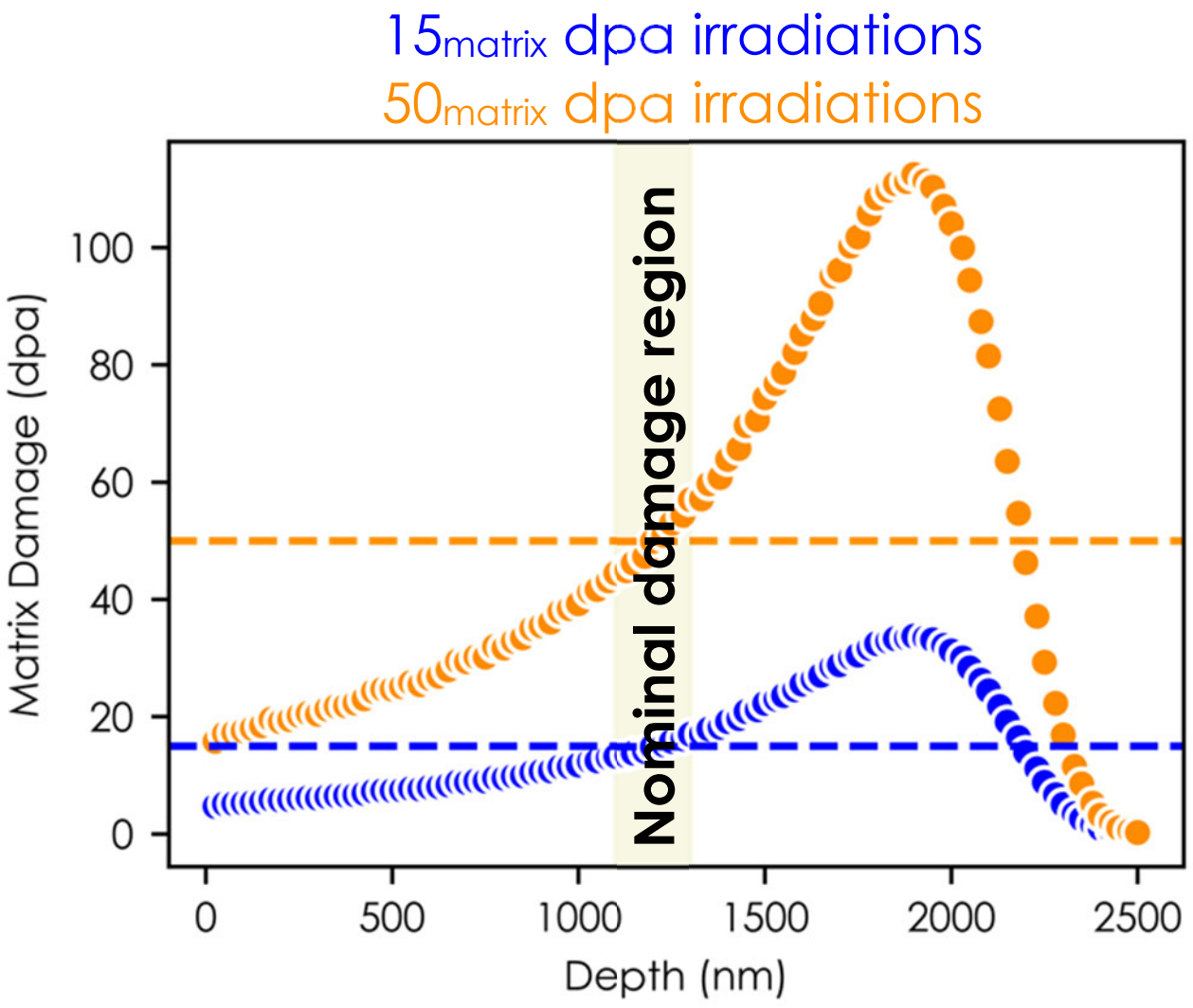

Figure SEQ Figure * ARABIC 2 Example damage profiles for the 15 (blue line) and 50 dpa (orange line) irradiations (ion direction is left-to-right of page). The target damage levels were reached in the nominal damage region 1,100-1,300 nm beneath the surface.

The average and standard error values of MX-TiC precipitate number density ($\rho$), equivalent diameter ($d$), and volume fraction ($f$) were calculated for each irradiation condition by assessing the MX-TiC precipitates in the STEM-EDS maps taken: the precipitates in each sample were counted to obtain $\rho$, the major ($a$) and minor ($b$) axes of each precipitate were measured to obtain the equivalent diameter ($d = \sqrt{a \times b}$), and the total volume of precipitates per condition was calculated to obtain volume fraction. The image processing software called FIJI, or ImageJ, was used for this analysis [55]. The standard errors reported represent the heterogeneity present in those variables. The Gaussian kernel density estimation with the built-in Python 3.7 Seaborn violin plot function [56] and with the built-in bandwidth Scott criterion [57] assessment was used to visualize the MX-TiC precipitate size distributions. The violin plots were scaled by width. Individual measurements of precipitate diameters were also overlaid onto the violin plots with symbols to illustrate the total precipitate number count for each condition studied.

## 1.3 Results and Discussion

### 1.3.1 As-received samples

The as-received (AR) sample was never thermally annealed after initial heat treatment or irradiated. An example STEM-EDS Ti map with the corresponding STEM-annular bright field (ABF) micrograph of the as-received condition of CNA9 are shown in Figure 3. Refer to Supplemental B for a full analysis of the precipitates in the as-received sample. The mean number density of $(2.3\pm0.3)\times10^{21}$ $m^{-3}$ MX-TiC precipitates in the as-received CNA9 sample is about 1-2 orders of magnitude greater than MX-(V,Nb)(C,N) precipitates found in Grade 91 steel, a second generation FM steel, and earlier generations of 9Cr RAFM steels [46,48]. The MX precipitates in the as-received sample had an average equivalent diameter of 7.8±0.3 nm, a maximum diameter of 13.1 nm, and a minimum diameter of 3.3 nm.

### 1.3.2 Temperature series to $7_{TiC}/15_{matrix}$ dpa

The first irradiation series tested CNA9 samples at 300, 400, 500, and 600ºC to $7_{TiC}/15_{matrix}$ dpa. However, it is necessary to benchmark the MX-TiC precipitate behavior in samples subjected only to thermal annealing to differentiate the effects of thermal annealing from irradiation at these temperatures. An assessment of the precipitates subjected only to thermal annealing at 300, 400, 500, and 600ºC for ~6 hours, which is the time of the irradiation to $7_{TiC}/15_{matrix}$ dpa with $7\times10^{-4}$ dpa/s, was conducted. Figure 4 shows the STEM-EDS Ti maps with the corresponding size distribution of precipitates for each annealing condition examined. These distributions are also compared to the as-received precipitate size distribution for reference. The dashed lines on the violin plots represent the 25% and 75% interquartile lines. Violin plots typically plot the median

Example microstructure of as-received sample of the size distribution but here the mean values of the diameters for each condition are shown in between the 25% and 75% interquartile lines.

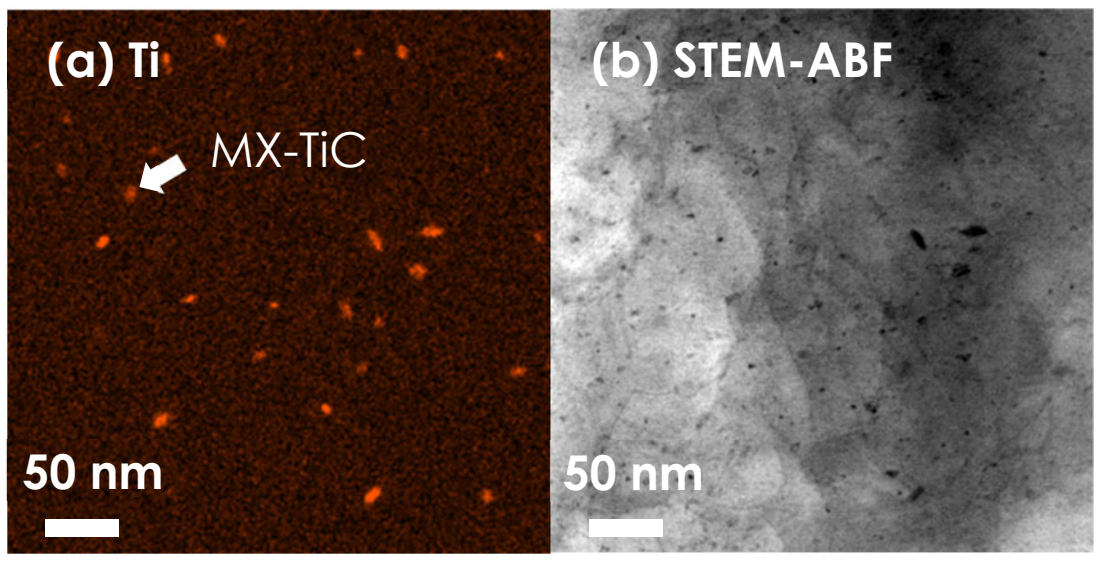

Figure SEQ Figure * ARABIC 3 An example (a) STEM-EDS Ti map with the corresponding (b) STEM-ABF micrograph of the as-received sample.

It was concluded that the precipitates remained thermally stable at the temperatures and time at temperatures tested as compared to the as-received condition. Accordingly, the evolution in precipitate statistics and size distributions will be ascertained by comparing the irradiated results to the results from the aggregated as-received and thermally annealed precipitate populations. This aggregated precipitate population is referred to as the control condition (CTRL) and is shown as the right-most size distribution in Figure 4. For the sake of clear visualization, the circular symbols representing individual counts of precipitates in the control condition will not be overlaid in subsequent figures, as the numerous counts obscure the shape of the violin plot.

To further assess the inherent heterogeneity present in the CNA9 microstructure, the ratios of number densities ($\rho/\rho_{CTRL}$), equivalent diameters ($d/d_{CTRL}$), and volume fractions ($f/f_{CTRL}$) of the precipitates present in the as-received and in each thermally annealed condition were taken as a function of the control condition. This statistical analysis is shown in the table beneath the precipitate size distributions in Figure 4. To determine the statistical significance of changes to precipitates caused by irradiation, the irradiated precipitate size distribution and statistics will be compared to the most conservative values of the ratios $\rho/\rho_{CTRL}$, $d/d_{CTRL}$, and $f/f_{CTRL}$ tabulated in Figure 4. These conservative values are shown in Table 4 for reference.

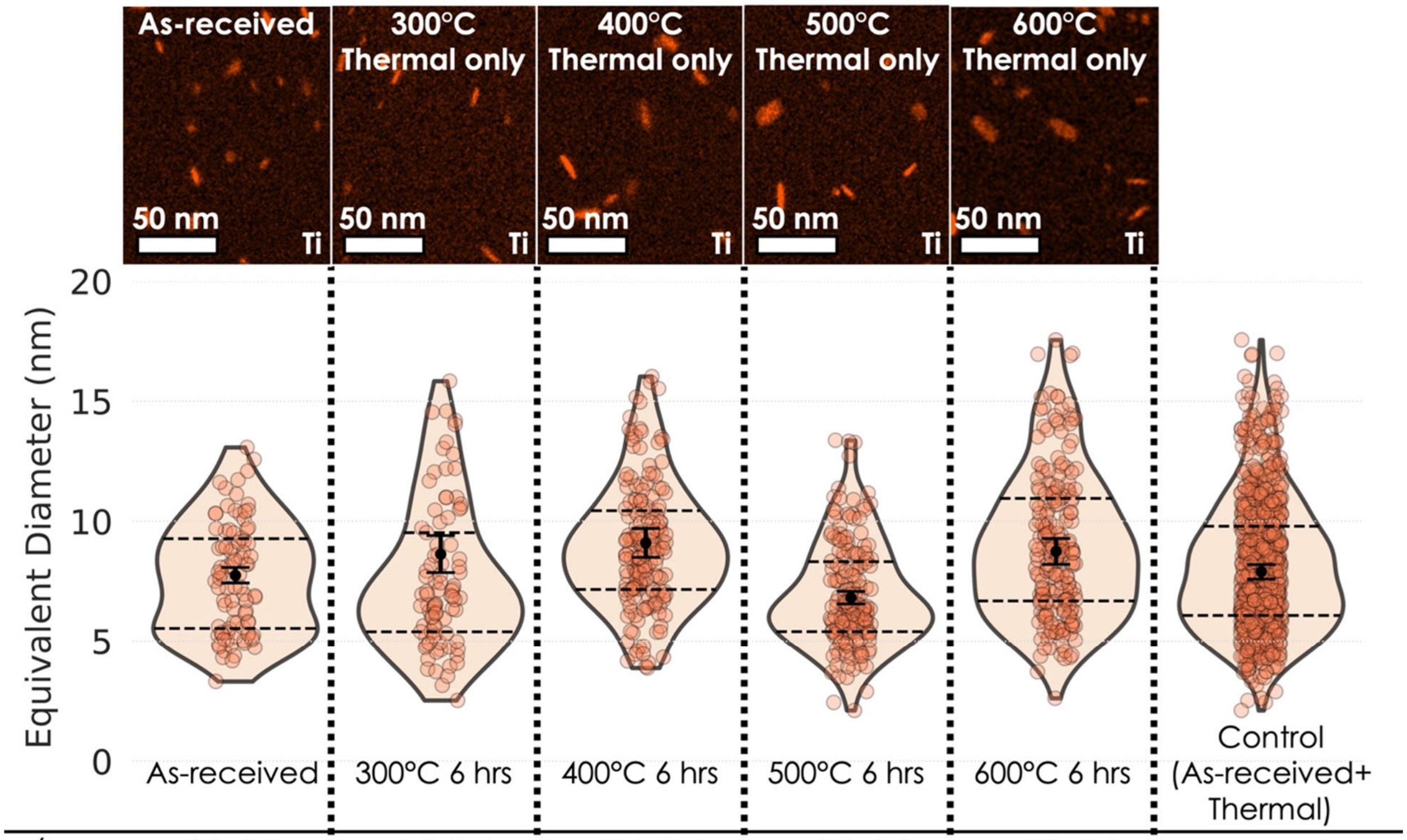

| | As-received | 300°C 6 hrs | 400°C 6 hrs | 500°C 6 hrs | 600°C 6 hrs | Control (As-received+ Thermal) |
|---|---|---|---|---|---|---|
| $\rho/\rho_{CTRL}$ | 0.8±0.1 | 1.0±0.4 | 1.1±0.4 | 0.9±0.2 | 1.3±0.4 | 1 |
| $d/d_{CTRL}$ | 1.0±0.1 | 1.1±0.1 | 1.2±0.1 | 0.9±0.1 | 1.1±0.1 | 1 |
| $f/f_{CTRL}$ | 0.8±0.1 | 1.0±0.3 | 1.4±0.3 | 0.6±0.2 | 1.8±0.7 | 1 |

Figure SEQ Figure * ARABIC 4 STEM-EDS micrographs with corresponding size distributions and precipitation statistics for the as-received specimen and the thermally annealed conditions at 300, 400, 500, and 600°C. The aggregated precipitate populations of the as-received and all thermally annealed conditions are shown via the distribution labeled 'Control.' Each condition was annealed for 6 hours. The dashed lines on the violin plots represent the 25% and 75% interquartile lines, in between which is the mean value of the diameter for that condition.

Table 4 Tabulated values from Figure 2 and Figure 3 that represent the range of statistical significance for ratio calculations between irradiated and control specimens.

| Ratio | Significant if less than | Significant if greater than |
|---|---|---|
| $\frac{\rho_{IRR}}{\rho_{CTRL}}$ | 0.6 | 1.7 |
| $\frac{d_{IRR}}{d_{CTRL}}$ | 0.8 | 1.3 |
| $\frac{f_{IRR}}{f_{CTRL}}$ | 0.4 | 2.5 |

Now that the general statistics of the precipitates in the control condition have been formulated, the precipitates present in the samples irradiated to $7_{TiC}/15_{matrix}$ dpa can be analyzed. Figure 5 shows the STEM-EDS micrographs for each condition in the temperature series (300, 400, 500, and 600ºC) irradiated to $7_{TiC}/15_{matrix}$ dpa along with any changes present from the control size distribution of the MX-TiC precipitates based on the split violin plots where the right side contains the irradiation-induced size distribution. The statistical precipitate responses to irradiation as compared to the control condition are displayed below the size distribution plots ($\rho_{IRR}/\rho_{CTRL}$, $d_{IRR}/d_{CTRL}$, and $f_{IRR}/f_{CTRL}$), where bolded values in the figure mean that variable underwent a statistically significant change with irradiation (refer to Table 4). Refer to Supplemental C for all data on precipitate statistics in each irradiated condition (*i.e.*, number of EDS maps taken, number of precipitates counted, etc.).

The first noticeable effect of irradiation is the significant decrease in the number density of MX-TiC precipitates at low temperatures (300 and 400°C), as shown by the reported values for $\rho_{IRR}/\rho_{CTRL}$: **0.3±0.1** and **0.5±0.1**, respectively (Figure 5, Table 4). However, the average precipitate sizes remained stable after irradiation ($d_{IRR}/d_{CTRL}$ = 1.1±0.1 and 1.0±0.1, respectively, for 300 and 400°C) as well as the 25% and 75% interquartile values. The resultant decreases in volume fractions are then primarily derived from the decreases in number densities ($f_{IRR}/f_{CTRL}$ = **0.4±0.1** and 0.5±0.1, respectively).

The most noticeable effect of irradiation at elevated temperatures is the upward shifts in the tail of the size distributions at 500 and 600°C. This is reflected in the significant increases in average equivalent diameters ($d_{IRR}/d_{CTRL}$ = **1.3±0.1** and **1.9±0.3**, respectively, for 500 and 600°C). The increases in precipitate sizes are most pronounced at 600°C, where the 25% interquartile line increased by ~45%, the mean increased by ~100%, the 75% interquartile line increased by ~110%, and the maximum increased by ~105% over the control condition. At 500°C, only the average size and the 75% interquartile line increased over the control condition. This difference may be due to the increased diffusion at 600°C causing greater levels of coarsening across the precipitate size spectrum. The number densities of precipitates also decreased significantly with irradiation at 500 and 600°C (**0.6±0.1** and **0.5±0.1**, respectively). As the volume fraction is most sensitive to changes in precipitate size ($f \sim r^3$), the volume fraction significantly increased at 600°C despite the decrease in number density (**4.7±1.4**).

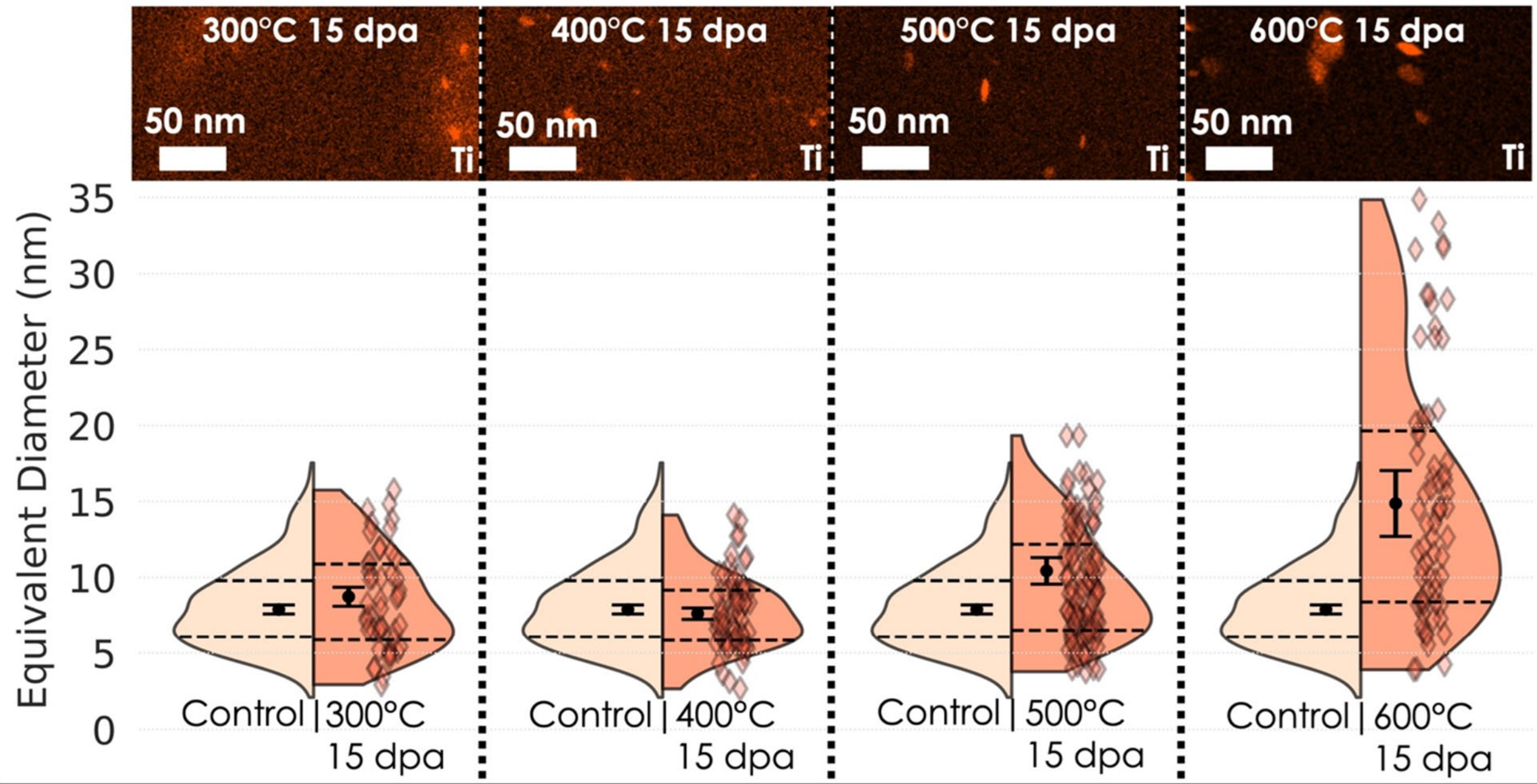

| | | | | |
|---|---|---|---|---|
| $\rho_{IRR}/\rho_{CTRL}$ | **0.3±0.1** | **0.5±0.1** | **0.6±0.1** | **0.5±0.1** |
| $d_{IRR}/d_{CTRL}$ | 1.1±0.1 | 1.0±0.1 | **1.3±0.1** | **1.9±0.3** |
| $f_{IRR}/f_{CTRL}$ | **0.4±0.1** | 0.5±0.1 | 1.2±0.3 | **4.7±1.4** |

Figure SEQ Figure * ARABIC 5 STEM-EDS micrographs for the control specimen and each irradiated condition in the single beam temperature series irradiated to $7_{TiC}/15_{matrix}$ dpa, along with the corresponding split violin plots and the statistical precipitate responses to irradiation ($\rho_{IRR}/\rho_{CTRL}$, $d_{IRR}/d_{CTRL}$, and $f_{IRR}/f_{CTRL}$). Statistically significant changes in the precipitation response to irradiation are bolded. Symbols depicting individually measured precipitates are not shown overlaid on the control size distribution for clarity, refer to Figure 4 for such a representation.

In summary, at lower temperatures (300 and 400°C), ballistic dissolution appears to be the dominant process, resulting in partial dissolution of MX-TiC precipitates by $7_{TiC}/15_{matrix}$ dpa. These results are in line with previous literature that shows ballistic dissolution to be more dominant at lower temperatures [58]. However, the precipitate size behavior at 500 and 600°C may be consistent with radiation-enhanced coarsening [59, 60]. An analysis provided no conclusive evidence that Ostwald-ripening was the primary coarsening mechanism at 500°C (Supplemental D). Even if radiation-induced Ostwald ripening is not occurring or is merely one of the coarsening mechanisms occurring, diffusion-mediated coarsening appears to be operational on precipitate stability at 500 and 600°C as shown by the upward shifts of the precipitate size distributions and significant increases in the average equivalent diameters over the control condition.

### 1.3.3 Damage level series at 500°C

Based on the previous set of experiments, precipitate behavior at a temperature of 500°C irradiated to $0.5_{TiC}/1_{matrix}$, $2.3_{TiC}/5_{matrix}$, $7_{TiC}/15_{matrix}$, $23_{TiC}/50_{matrix}$, and $47_{TiC}/100_{matrix}$ dpa were chosen for further evaluation. The temperature of 500°C was chosen as it represents a shift in precipitate behavior from ballistic dissolution at lower temperatures to radiation-enhanced growth; noticeable coarsening took place at $7_{TiC}/15_{matrix}$ dpa and it is of interest to determine if this coarsening is a function of damage level; and 500°C is most likely closest to the peak swelling temperature of CNA9, which will be important for ascertaining precipitate behavior-swelling relationships in future corresponding dual ion irradiations which were completed as a part of this series of works.

Figure 6 highlights the significant impact of damage level on precipitate dissolution. The precipitates do not undergo statistically significant changes at low damage levels in density, size, or volume fraction – essentially, they are stable at and below 5 $dpa_{matrix}$. As discussed in the previous section, precipitates at 15 $dpa_{matrix}$ displayed radiation-enhanced growth. However, the MX-TiC precipitates have completely dissolved at 50 $dpa_{matrix}$ and 100 $dpa_{matrix}$. These experiments show that the mechanisms of precipitate behavior at 500°C is dependent on damage

level – displaying an evolution of stability ($0.5_{TiC}/1_{matrix}$, $2.3_{TiC}/5_{matrix}$ dpa), growth ($7_{TiC}/15_{matrix}$ pa), and dissolution ($23_{TiC}/50_{matrix}$, $47_{TiC}/100_{matrix}$ dpa).

### 1.3.4 Temperature series to $23_{TiC}/50_{matrix}$ dpa

The MX-TiC precipitates were shown to behave differently at low versus high temperatures at the intermediate damage level of $7_{TiC}/15_{matrix}$ dpa in previous sections. To further explore the temperature dependencies, two more irradiations at 300 and 500°C were conducted to high damage level. These temperatures were chosen because precipitates were shown to be in different regimes of stability at 300 and 500°C (partial dissolution and radiation-enhanced coarsening, respectively) at the intermediate damage level. Thus, this experimental series explores the evolution of those mechanisms at the elevated damage level of $23_{TiC}/50_{matrix}$ dpa.

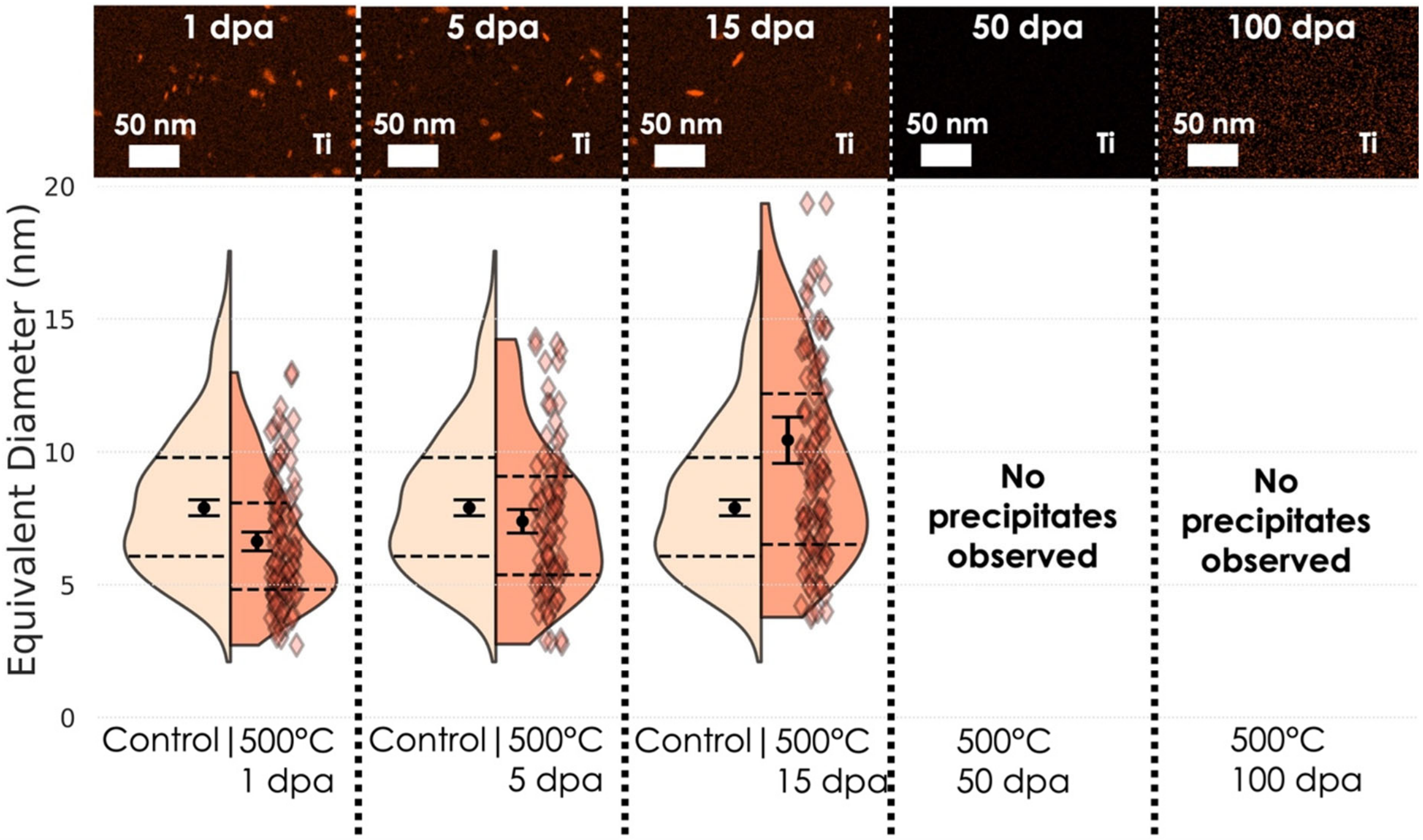

| | | | | | |
|---|---|---|---|---|---|
| $\rho_{IRR}/\rho_{CTRL}$ | 1.2±0.4 | 1.1±0.4 | **0.6±0.1** | | |
| $d_{IRR}/d_{CTRL}$ | 0.8±0.1 | 0.9±0.1 | **1.3±0.1** | | |
| $f_{IRR}/f_{CTRL}$ | 0.8±0.2 | 1.0±0.4 | 1.2±0.3 | | |

Figure SEQ Figure * ARABIC 6 STEM-EDS micrographs of Ti with corresponding MX-TiC precipitate size distributions and statistics for the control specimen and specimens irradiated at 500°C to 1, 5, 15, 50, and 100 $dpa_{matrix}$. Individual measurements of precipitates are shown by markers overlaid on the violin plot size distributions.

Example STEM-EDS composition maps of Ti in Figure 6 display that the MX-TiC precipitates dissolved by $23_{TiC}/50_{matrix}$ dpa at both 300 and 500°C in the nominal irradiation region. It is possible that not all the MX-TiC precipitates dissolved in these conditions, but that the precipitate density was below detection due to the limited volume able to be assessed with the STEM-EDS technique or that the precipitate sizes were below the STEM-EDS resolution of ~0.5 nm. As no precipitates were present in the irradiated conditions, a detailed analysis of the precipitates in the thermally annealed conditions was not conducted. However, upon inspection of STEM-EDS maps taken of thermally annealed precipitates for these conditions, the precipitate sizes were visually in line with the control condition, though the number density appeared to decrease in the 500°C thermal condition (Supplemental E). MX-TiC precipitates display a high level of thermal stability in the literature as well [12, 30]. These experiments show that damage level is the dominant factor in precipitate dissolution by $23_{TiC}/50_{matrix}$ dpa for the temperatures (300, 500°C) and damage rate ($7\times10^{-4}$ dpa/s) studied.

### 1.3.5 Precipitate Stability Model

Thus far, the MX-TiC precipitates have been experimentally shown to display temperature-dominated responses at intermediate damage level ($7_{TiC}/15_{matrix}$ dpa) and damage-dominated responses by high damage level ($23_{TiC}/50_{matrix}$ dpa). To further explore these precipitate dependencies, the Muroga, Kitajima, and Ishino (MKI) model of precipitate stability was employed to the irradiations conducted at 300 and 500°C to $7_{TiC}/15_{matrix}$ and $23_{TiC}/50_{matrix}$ dpa to assess the effects of temperature and damage level [61].

The MKI model calculates the precipitate size and dissolution under irradiation using the framework of recoil resolution [62]. Recoil resolution is also referred to as ballistic dissolution in literature, where collision cascades during irradiation cause the ballistic ejection of solute atoms from precipitates into the matrix [58]. The MKI model assumes a perfectly homogeneous unimodal distribution of spherical particles pre-existing before irradiation. This was not the case in CNA9 (Figure 4) and as such may affect the applicability of the model. In addition, the model does not capture the physical mechanistic effects of coarsening of an initially polydisperse distribution of precipitates [60]. As the model calculates precipitate behavior as a function of damage level, the method employed to calculate damage level (*i.e.*, conventional quick-KP SRIM calculations or perhaps correcting calculations to account for the athermal recombination of point defects (arcdpa) and atomic mixing (replacements per atoms, rpa) [53]) will affect the results such as the critical dose to dissolution. However, it is assumed that the general conclusion drawn from the MKI model here will remain the same, despite if the exact magnitude of certain effects as a function of dose may be different than those found based on experimental results. It is assumed that the balance between diffusional and displacement effects will still evolve as a function of damage level and variability between the interplay of these two effects will still impact the growth and dissolution prediction of precipitates.

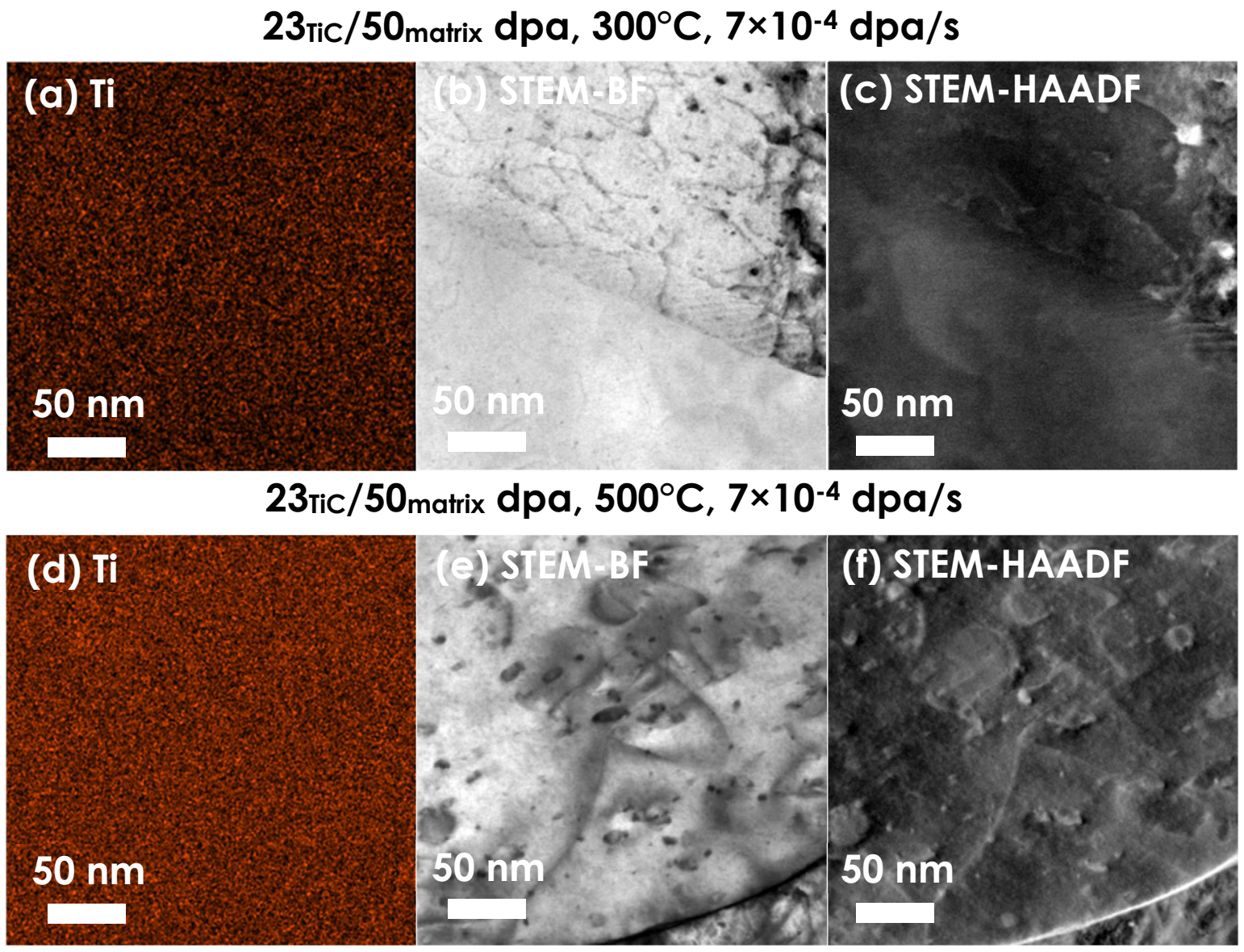

Figure SEQ Figure * ARABIC 7 (a) STEM-EDS Ti map encompassing the nominal damage region with corresponding (b) STEM-BF and (c) STEM-HAADF maps for the sample irradiated to $23_{TiC}/50_{matrix}$ dpa at 300°C. (d) STEM-EDS Ti map encompassing the nominal damage region with corresponding (e) STEM-BF and (f) STEM-HAADF maps for the sample irradiated to $23_{TiC}/50_{matrix}$ dpa at 500°C.

The MKI model introduces a variable known as the recoil resolution efficiency, $\varepsilon_{res}$, which is the probability of dissolution under irradiation. It is defined as the number of atoms moved from inside to outside the precipitate with collision cascades divided by the number of atomic displacements within the precipitate:

$$\varepsilon_{RES} = \frac{12\left(\frac{R_{eff}}{r_p}\right) - \left(\frac{R_{eff}}{r_p}\right)^3}{16} \qquad \text{Eq. 1}$$

where $r_p$ is the precipitate radius and $R_{eff}$ is the effective range of recoiling solute atoms from precipitates into the matrix. $R_{eff}$ modifies the physical recoil range of Ti solute atoms, R, by the probability of those solute atoms to return to the original precipitate, $r_{cap}$. The physical range of Ti atoms, R, that recoil from the precipitates to the matrix was calculated to be 0.7 nm, determined with the use of SRIM (see Supplemental F for details on this calculation). The likelihood of solute atoms to diffuse back to the precipitate from the matrix is referred to as the capture radius, $r_{cap}$. $r_{cap}$ serves as a mechanism to incorporate solute and defect migration in the diffusion-less MKI model.

This work has defined a new relationship to calculate $R_{eff}$ from R and $r_{cap}$, graphically shown in Figure 7:

$$R_{eff} = R - r_{cap} \qquad \text{Eq. 2}$$

A condition of stability is met if $R = r_{cap}$. The recoiled Ti solute atoms remain in the matrix if $R > r_{cap}$ (*i.e.*, precipitates tend toward dissolution), resulting in positive values of $R_{eff}$ and $\varepsilon_{res}$. The recoiled Ti solute atoms migrate to the original precipitate or a neighboring precipitate if $R < r_{cap}$. (*i.e.*, precipitates tend toward growth), resulting in negative values of $R_{eff}$ and $\varepsilon_{res}$.

Quantitative calculations of $R_{eff}$ rest on having a value for $r_{cap}$ to input into Eq. 2. Though the factors that affect $r_{cap}$ are known (*i.e.*, precipitate stress fields inducing preferential defect drifts, temperature affecting the diffusion of solutes to precipitates), there is no current methodology in literature known to the authors at the time of this work to quantitively calculate $r_{cap}$ as a function of all the known factors. Instead, this work will use physically relevant values of $R_{eff}$ that will survey the influence of small, intermediate, and large $r_{cap}$ values on precipitate behavior. The results of this analysis will develop an understanding of the relationship between factors that decrease or increase $r_{cap}$ with general precipitate behavior. Thus, the following values of $R_{eff}$ were chosen to sample the effect of $r_{cap}$ on precipitates: -0.8, 0.05, and 0.7 nm, which correspond to comparatively large, intermediate, and negligible values of $r_{cap}$.

The precipitate size as a function of damage level can be calculated by integrating the rate of the precipitate size change under irradiation, ensuring that $\varepsilon_{res}$ is a function of $r_p$:

$$\int_{t_0}^{t_1} \frac{dr_p}{dt} = \int_{r_0}^{r_1} \frac{-r_p S \varepsilon_{RES}}{3} \qquad \text{Eq. 3}$$

where $r_0$ is the initial average precipitate radius in the control condition at 0 dpa, $r_1$ is the calculated radius value up to $23_{TiC}/50_{matrix}$ dpa, and $t_1$ is the time it took to reach $23_{TiC}/50_{matrix}$ dpa. The solve_ivp function in the SciPy package (version 1.6.2) in Python [63] was used to integrate Eq.

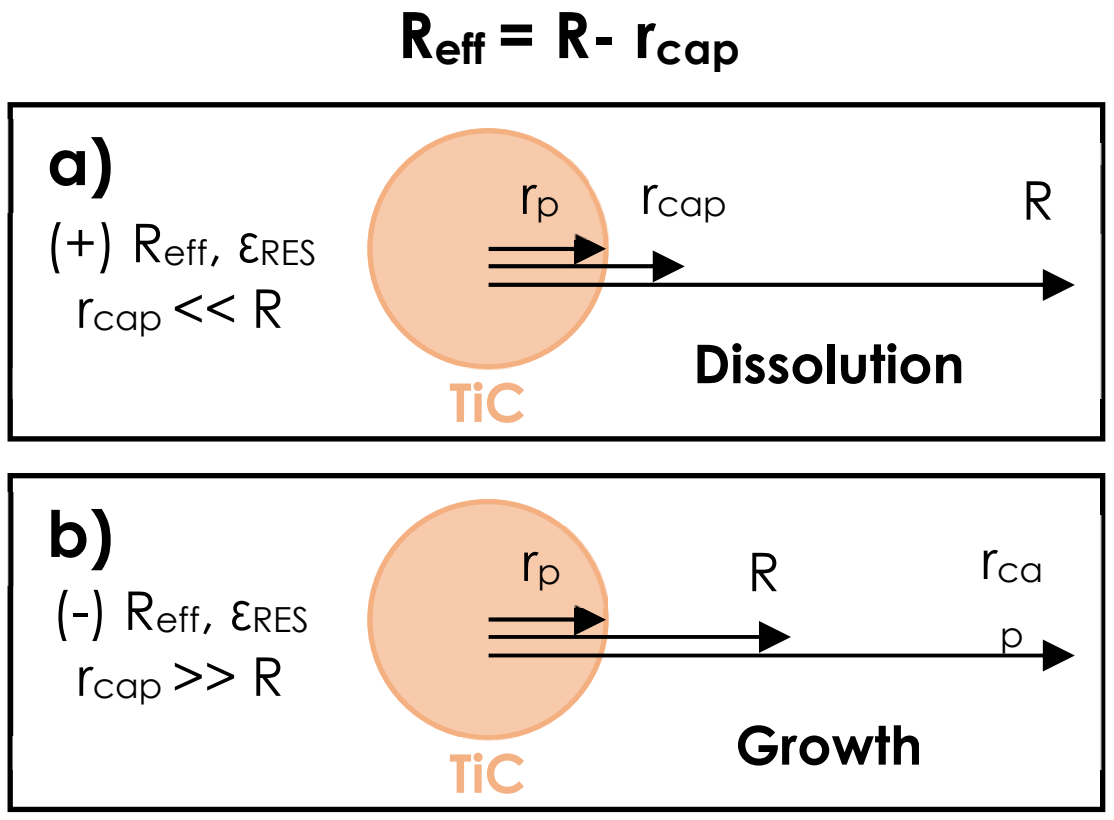

Figure SEQ Figure * ARABIC 8 Pictorial representation of the relationship between $R_{eff}$, R, and $r_{cap}$ where $R_{eff}$ = R- $r_{cap}$. (a) The capture radius, $r_{cap}$, is less than the physical recoil range, R, resulting in a positive $R_{eff}$ and $\varepsilon_{res}$. As such, more solute atoms leave the precipitate than are replenished during irradiation and the precipitates tend toward dissolution. (b) $r_{cap}$ is greater than R, resulting in a negative $R_{eff}$ and $\varepsilon_{res}$. As such, more solute atoms enter the precipitate than leave during irradiation and the precipitates tend toward growth.

3.

Figure 9 displays the results of this analysis, where Figure 9a shows the results for 300°C and Figure 9b for 500°C. The y-axis is the precipitate radius as a function of damage level and the x-axis is the MX-TiC damage level. The white boxplot in each subplot at 0 dpa represents the distribution of the control specimen. The irradiated experimental precipitate size distributions are shown with orange box plots at $7_{TiC}/15_{matrix}$ dpa. The dissolution by $23_{TiC}/50_{matrix}$ dpa at 300 and 500°C are shown with square markers. The black dotted line represents the evolution of $r_p$ with $R_{eff}$ = 0.05 nm, the gray solid lines represent $R_{eff}$ = 0.8 nm, and the gray dotted lines represent $R_{eff}$ = -0.8 nm.

For the 300°C irradiation (Figure 9a), the value of $R_{eff}$ = 0.05 nm (black dashed lines) provides the most accurate simulation of the experimentally observed average precipitate size $7_{TiC}/15_{matrix}$ dpa. The value of $R_{eff}$ = 0.05 nm corresponds to an intermediate, positive value of $r_{cap}$ and to a small, positive value of $\varepsilon_{res}$, translating to slightly more solutes leaving the precipitate than are replenished. As such, partial dissolution of precipitates would be expected from the model. This is in line with experimental observations of precipitates at 300°C to $7_{TiC}/15_{matrix}$ dpa where it was found that the irradiated precipitates did not display statistically significant changes in size distribution from the control specimen but did display statistically significant dissolution. At this condition, $r_{cap}$ is theorized to be primarily influenced by the low diffusion of solute atoms back to the precipitate resulting from the low temperature, leading to partial dissolution.

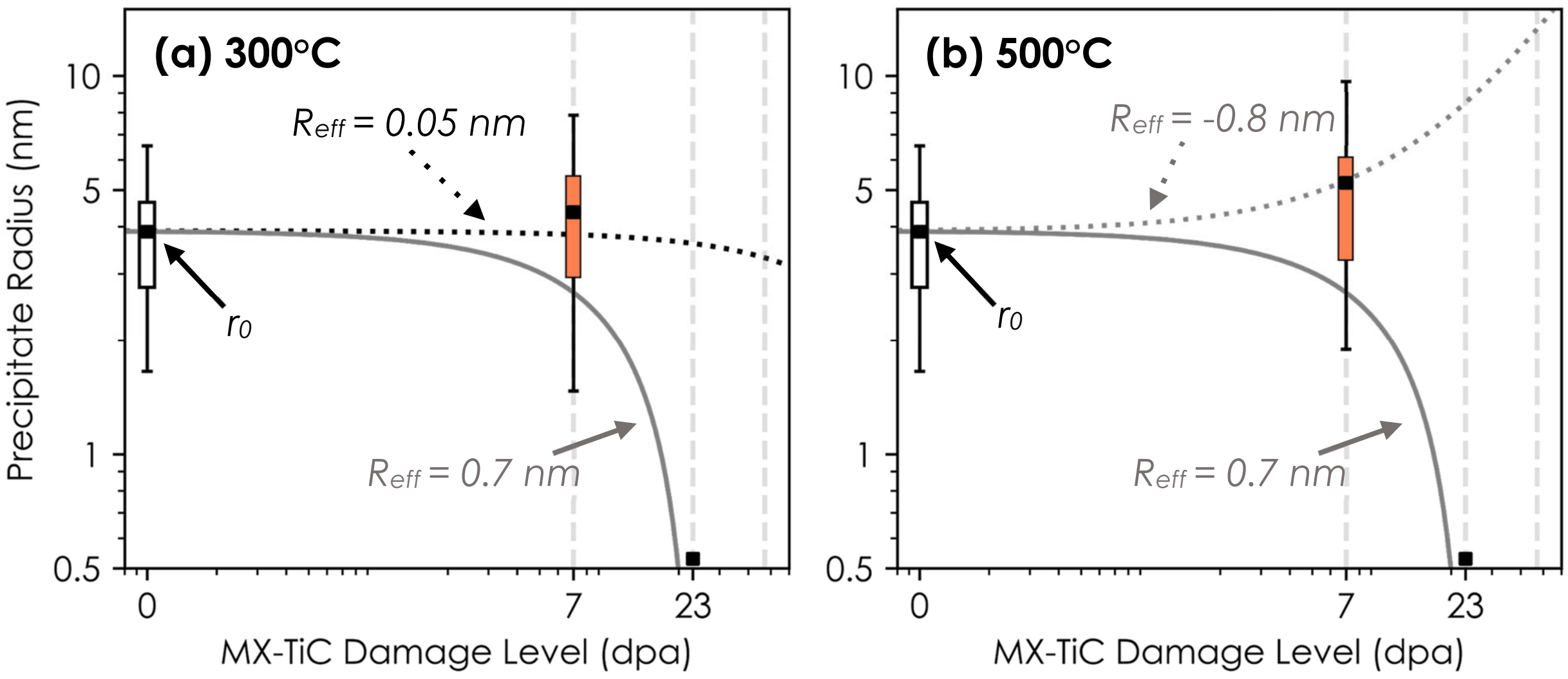

Figure SEQ Figure * ARABIC 9 Plots showing the theoretical behavior of MX-TiC precipitates under irradiation as modelled by the MKI
model of recoil resolution for (a) 300°C and (b) 500°C as a function of various values of $R_{eff}$. The white box plot at 0 dpa in each plot shows the size distribution of the control specimen. The orange box plots shown on each plot represent the experimentally observed precipitate size distribution at $7_{TiC}/15_{matrix}$ dpa for both temperatures and the square markers represent the dissolution of precipitates by $23_{TiC}/50_{matrix}$ dpa. The x-axis represents the damage level received by the MX-TiC precipitates.

For the high temperature irradiation at 500°C to $7_{TiC}/15_{matrix}$ dpa, a value of $R_{eff}$ = -0.8 nm (dotted black lines) accurately captures the coarsening trend for this condition (Figure 9b). This value of $R_{eff}$ corresponds to a value of $r_{cap}$ that is greater than R and to a large, negative value of $\varepsilon_{res}$, which matches the experimentally observed coarsening. The elevated temperature of 500°C is theorized to cause greater diffusion of solute atoms as compared to the 300°C condition, thereby increasing the value of $r_{cap}$ and leading to precipitate growth.

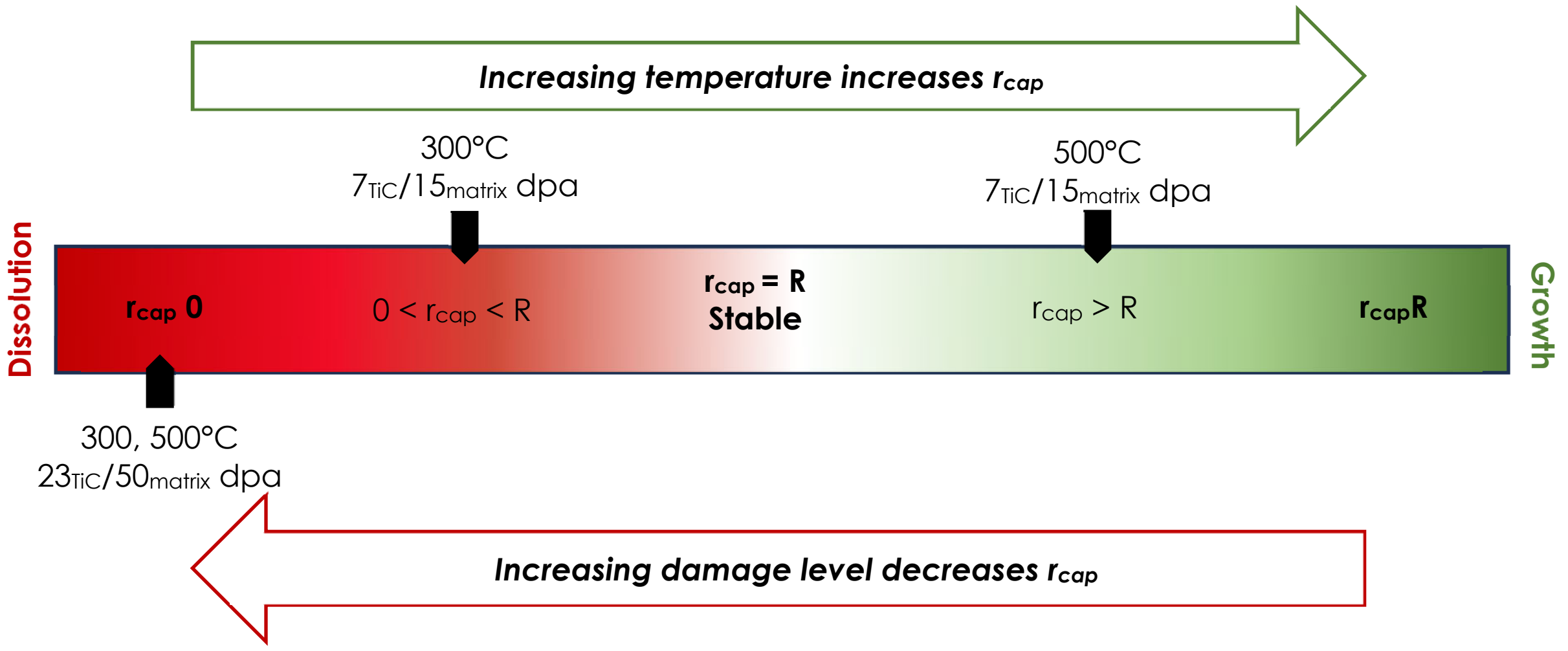

Figure SEQ Figure * ARABIC 10 Graphical representation of how $r_{cap}$ changes as a function of temperature and damage level: $r_{cap}$ increases with increasing temperature, leading to precipitate growth, and $r_{cap}$ decreases with increasing damage level, leading to dissolution. The values of $r_{cap}$ directly affect $\varepsilon_{res}$.

Up to this point, the values of $r_{cap}$ assessed at intermediate damage level do not account for the complete dissolution of precipitates by $23_{TiC}/50_{matrix}$ dpa at 300 and 500°C. In order to simulate the dissolution of precipitates at the high damage level conditions, a large value of $R_{eff}$, corresponding to a negligible value of $r_{cap}$, is needed. As such, the value of $R_{eff}$ = 0.7 nm (solid gray lines) displays the proper dissolution behavior by $23_{TiC}/50_{matrix}$ dpa. This analysis suggests that the dominant influence on precipitate behavior switches from temperature at intermediate damage level to dissolution at high damage level – in other words, $r_{cap}$ is a function of damage level. These results are summarized in Figure 10. The next section will discuss the possible reasons why $r_{cap}$ and $\varepsilon_{res}$ change as a function of damage level.

### 1.3.6 Factors affecting precipitate stability as a function of damage level

First, $r_{cap}$ may have been influenced by the interfacial dislocation structure at the semi-coherent MX precipitate-matrix interface (Supplemental G). The strain field around a precipitate, and thus point defect diffusivity, is partially controlled by the interfacial dislocation structure [64, 65]. In addition, previous studies show that as damage levels increase, interfacial dislocation structures become unstable, leading to changes in precipitate size or to precipitate dissolution [66-68]. Thus, the diminishing strains around the MX precipitates and the loss of interfacial dislocation structures that may have occurred as the damage level increased past $7_{TiC}/15_{matrix}$ dpa could have potentially led to changes in solute replenishment or the alteration of the precipitate defect bias, leading to dissolution. Such interfacial changes would have decreased $r_{cap}$ with increasing damage level.

Second, previous research on MX-TiC precipitates in austenitic Ti-modified stainless steel concluded that the dissolution of MX-TiC precipitates between 10 and 45 dpa corresponded to the formation of other Ti-containing secondary phases, such as phosphides and the G phase [69]. Thus, the Ti atoms redistributed from the MX precipitates to other features, causing their dissolution. Though the radiation-induced nucleation of new Ti-containing phases was not explicitly observed for CNA9, it is possible that Ti atoms redistributed to other features (*i.e.*, to the larger MX TiC precipitates or to grain boundaries) during irradiation which impacted the TiC solubility limit. Such a driving force for redistribution would have caused the diffusion of Ti atoms away from the MX-TiC precipitates, effectively causing $r_{cap}$ to approach zero.

Third, and related to Ti solute redistribution, is the interparticle spacing of the MX-TiC precipitates, *L*. Table 4 shows the interparticle spacing of precipitates and the mean distance traveled by a point defect before being captured by an MX-TiC precipitate ($k_{MX}^{-1}$), which is related to $L$, or any microstructural feature ($k_{tot}^{-1}$):

$$L = \frac{1}{2\sqrt{\rho d}} \text{ [70]} \quad \text{Eq. 4}$$

$$k_{MX}^{-1} = \frac{1}{\sqrt{k_{MX}^{2}}} \quad \text{Eq. 5}$$

$$k_{tot}^{-1} = \frac{1}{\sqrt{k_{tot}^{2}}} \quad \text{Eq. 6}$$

where $k_{MX}^{2}$ and $k_{tot}^{2}$ are the precipitate and total CNA9 sink strengths, respectively (Supplemental H). The smaller values of $k_{tot}^{2}$ versus $k_{MX}^{2}$ means that point defects are more likely to be captured by features other than MX precipitates. Thus, a high average interparticle spacing, as was found in CNA9, correlates to a decreasing $r_{cap}$ as recoiling Ti solutes are less likely to reach neighboring precipitates. A caveat to this analysis is that the interparticle spacing used is based on a spatially-averaged density of precipitates, when in truth the precipitates are heterogeneous (refer to Supplemental B for analysis of various STEM-EDS as-received maps). Though the localization of MX-TiC precipitates may affect local dissolution behavior, no such evidence could be found experimentally and the averaged interparticle spacing is assumed acceptable for this analysis. This analysis indicates that interparticle spacing may have an outsized effect on precipitate stability, potentially deriving from the ability of the MX precipitates to exchange Ti solutes during irradiation [71]. The dissolution effects of low volume fraction and thus high interparticle spacing was predicted by the original recoil resolution model, of which the MKI model is based on [62, 72].

In summary, it is assumed that the three aforementioned factors possibly affecting $r_{cap}$ operated concurrently to cause the dissolution of precipitates at high damage levels: the loss of initial precipitate interfacial characteristics, Ti solute redistribution, and the lack of exchange Ti solutes between MX precipitates derived from a high interparticle spacing. Table 5 provides a summary of the discussion on the recoil resolution model of MX-TiC precipitates in CNA9. Subsequent investigations could untangle the individual impacts of these three effects on precipitate stability. In addition, the role of C in the MX-TiC precipitate stability was not considered but could be a factor as well. As noted, the MKI model does not consider different-sized particles, but particle sizes are a dominant driving force for particle 440 coarsening and thus may also affect $r_{cap}$.

Notably, it is uncertain if the above analysis and conclusions will be applicable to the various types of MX precipitates present in RAFM steels because the morphological and compositional differences between the precipitates may lead to different interfacial characteristics, such as interfacial energy and misfitting dislocation structures, lattice strains, coarsening rates, solute recoil distances, interparticle spacing, binding between the constituent atoms of the precipitate, existence of other precipitate and phases, and solute solubility limits under irradiation. This field of research is ripe for developing quantitative understandings of precipitate evolution based on the listed interfacial characteristics, ultimately allowing for calculation of $r_{cap}$ as a function of damage level. Additionally, the ion irradiation experiments did not concurrently test the effects of creep, stress, and fatigue, which will be important for understanding how precipitates will respond under load during fusion plant operation. Lastly, there remains the open question of if helium co-injection during self-ion irradiation will alter the precipitate behavior. As helium is expected to be an important microstructural modifier under fusion operation [73], such research is needed. This is the topic of the next effort in this series of work by the authors.

Table 5 Mean MX interparticle spacing, $L$, and mean distance traveled by a point defect before being captured by a MX precipitate ($k_{MX}^{-1}$) or any microstructural feature ($k_{tot}^{-1}$).

| Condition | $L$ | $k_{MX}^{-1}$ | $k_{tot}^{-1}$ |
|---|---|---|---|
| 300°C to $7_{TiC}/15_{matrix}$ dpa | 180 nm | 144 nm | 18 nm |
| 500°C to $7_{TiC}/15_{matrix}$ dpa | 118 nm | 94 nm | 52 nm |

Table 6 Summary of the precipitate behavior in various temperature and damage level regimes from the experimental and MKI modeling results.

| **Damage Level Regime** | **Temperature Regime** | **Suggested dominant mechanism affecting $r_{cap}$** | **Precipitate behavior** | **Capture Radius ($r_{cap}$)** | **Effective range ($R_{eff}$)** | **Recoil resolution efficiency ($\varepsilon_{RES}$)** |
|---|---|---|---|---|---|---|
| Intermediate ($7_{TiC}/15_{matrix}$ dpa) | Low temperature (300, 400°C) | Low diffusion of Ti causing less solute migration back to precipitates | No size changes, but partial dissolution | Intermediate ($0 < r_{cap} < R$) | Small, positive | Small, positive |
| | High temperature (500, 600°C) | High diffusion of Ti causing more solute migration to precipitates | Precipitate growth | Large ($r_{cap} > R$) | Large, negative | Large, negative |
| High ($23_{TiC}/50_{matrix}$ dpa ) | All temperatures (300, 500°C) | Ti solute redistribution, loss of initial precipitate interfacial characteristics, and high interparticle spacing | Complete dissolution | Minimum ($r_{cap} = 0$) | Large, positive | Large, positive |

## 1.4 Conclusion

In conclusion, an advanced Fe-9Cr RAFM steel was self-ion irradiated to ascertain the effects of temperature and damage level on MX-TiC precipitate stability:

- At intermediate damage levels ($7_{TiC}/15_{matrix}$ dpa), MX-TiC precipitates displayed temperature-dependent behavior. Radiation-enhanced diffusion was the primary mechanism at elevated temperatures (500, 600°C), leading to precipitate coarsening, while ballistic dissolution was the primary mechanism at low temperatures (300, 400°C).

- The precipitate response was temperature insensitive at higher damage levels ($23_{TiC}/50_{matrix}$ dpa and $47_{TiC}/100_{matrix}$ dpa). At all temperatures tested at higher damage levels (300, 500°C), the MX precipitates fully dissolved.
- The temperature- and dose-dependent behavior of MX precipitates was modeled with the MKI model, which uses the recoil resolution model of precipitate stability under irradiation. With the use of the modeling and experimental results, it was theorized that temperature's effects on diffusion explained the coarsening and dissolution at intermediate damage levels, but that a complicated interplay of different factors caused the dissolution at higher damage levels. The factors included the loss of initial precipitate interfacial characteristics, Ti solute redistribution, and the lack of exchange Ti solutes between MX precipitates derived from a high interparticle spacing.
- Due to their dissolution by $23_{TiC}/50_{matrix}$ dpa, any potential advantages in terms of mechanical properties or resistance to radiation damage provided by these precipitates would be lost in the early stages of fusion power plant operation.

The effect of helium co-implantation on precipitate behavior will be assessed in the next report of this series by the authors.

**Declaration of competing interest**

The authors declare that they have no known competing financial interests or personal relationships that could have appeared to influence the work reported in this paper.

**Acknowledgements**

The experimental work presented here was funded by the Fusion Energy Sciences program (DOE-FOA-0002173). The authors also acknowledge the University of Michigan-Ann Arbor College of Engineering for financial support and the Michigan Center for Materials Characterization for use of the instruments and staff assistance. Research presented here was also partially supported by the Laboratory Directed Research and Development program of Los Alamos National Laboratory under project number XXPV. This research was partly sponsored by the US Department of Energy, Office of Fusion Energy Sciences under contract DE-AC05-00OR22725 with UT-Battelle, LLC.

**Notes**

Portions of this supplemental are reproduced from the dissertation of T.M. Kelsy Green [1] and are provided within for clarity and completeness to the reader and to mitigate possible access issues to the dissertation document hosted and supported by the University of Michigan-Ann Arbor.

**Supplemental A: Determination of Recoil Rate (dpa/s) of TiC Precipitates Under Irradiation**

An analysis was completed to convert the rate of damage in TiC precipitates in the nominal damage region when CNA9 was single ion irradiated with matrix damage rate of $7\times10^{-4}$ dpa/s. A quick KP calculation in SRIM was run with 5,000 ions using the base CNA9 composition provided within the parent work. The simulated layers were as follows: 1,200 nm of CNA9; 20 nm of TiC; and 1,280 nm of CNA9. Figure S.1 shows the SRIM damage results. The star on Figure S.1 notes the damage level for the TiC layer at ~7 dpa. The damage level for CNA9 is ~15 dpa, translating to a 0.467 difference between the damage levels of the TiC precipitates and the CNA9 matrix within in nominal damage region. The result is it is assumed that when the matrix had a damage rate of $7\times10^{-4}$ dpa/s the MX-TiC precipitates were exposed to a damage rate of $3.3\times10^{-4}$ dpa/s. Figure 2 in the main text shows the SRIM calculation without the TiC layer.

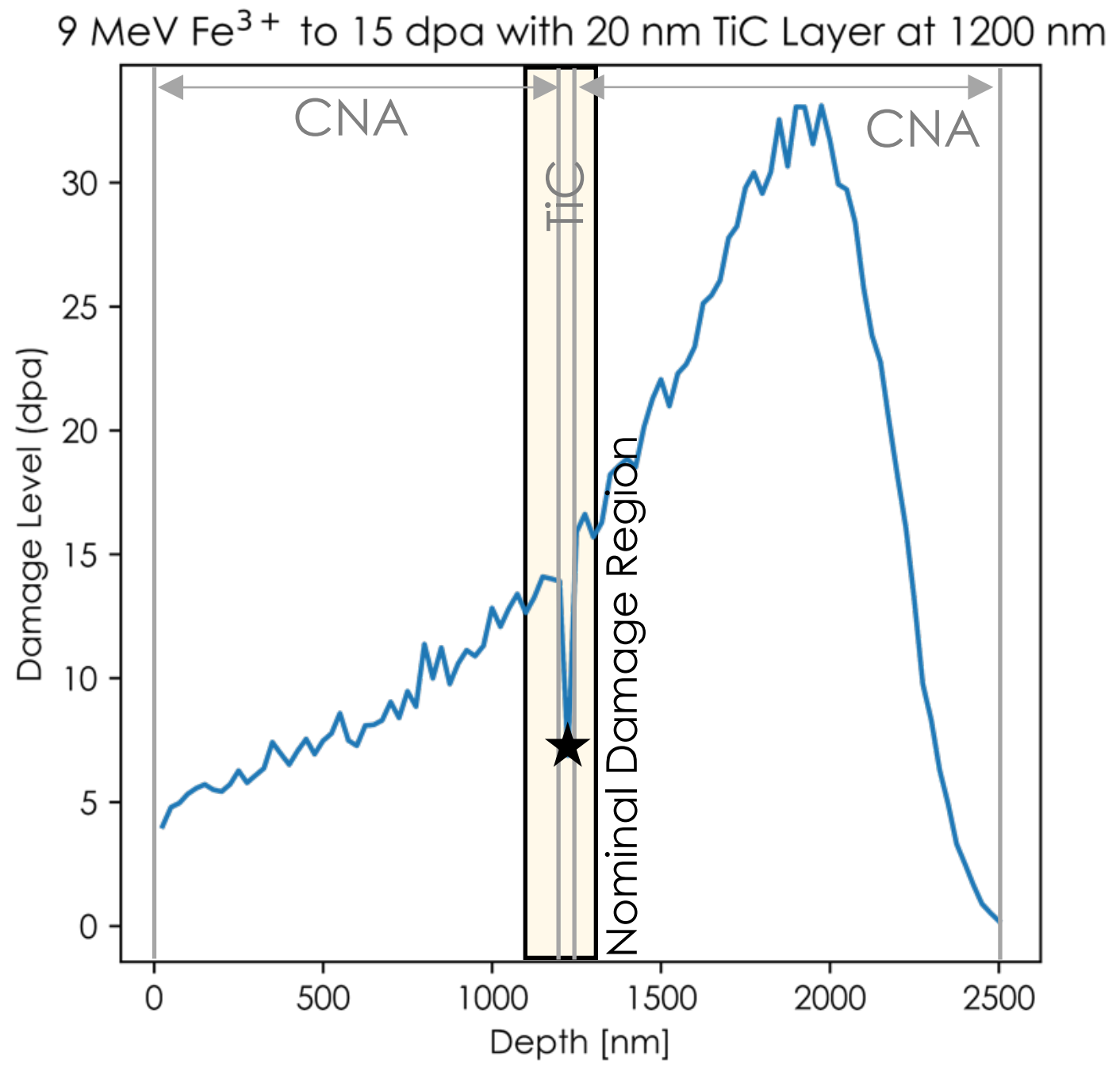

Figure S.1 SRIM calculation of damage into CNA9 with a layer of Ti and C at 1200 nm, simulating the damage that would occur in a TiC precipitate (star).

**Supplemental B: As-received Sample**

---

The as-received sample was provided by Oak Ridge National Laboratory and had no additional environmental exposures after the initial heat treatments outlined in the main body of this work. The STEM-EDS micrographs and corresponding STEM-BF images of the as-received specimen are shown in Figure S.2. shows the subsequent quantitive analysis including the size distributions and statistics from all STEM-EDS micrographs taken as part of this work. Figure S.3 shows the number of precipitates counted per map (*N*), number density of precipitates (*ρ*), average equivalent diameter (*d*), and volume fraction (*f*) for each map. Standard errors are used to indicate the possible counting errors associated with the analysis applied. The size distribution labeled 'Total' was created from all precipitates counted in the as-received sample from all four maps. The circular markers on Figure S.3 are the individual precipitate diameter measurements. . The dashed lines on the violin plots represent the 25% and 75% interquartile lines, in between which is the mean value of the diameter for that condition.

To analyze and measure the variation within the original sample the number density, size, and volume fraction of precipitates of each individual STEM-EDS map of the as-received specimen were contrasted with those calculated from the average of all maps (shown in the 'Total' column in Figure S.3). The resulting ratios are depicted in Figure S.3, denoted as $\rho/\rho_{CTRL}$ $d/d_{CTRL}$, and $f/f_{CTRL}$, and ratios then indicate the extent of heterogeneity in the initial microstructure prior to irradiation.

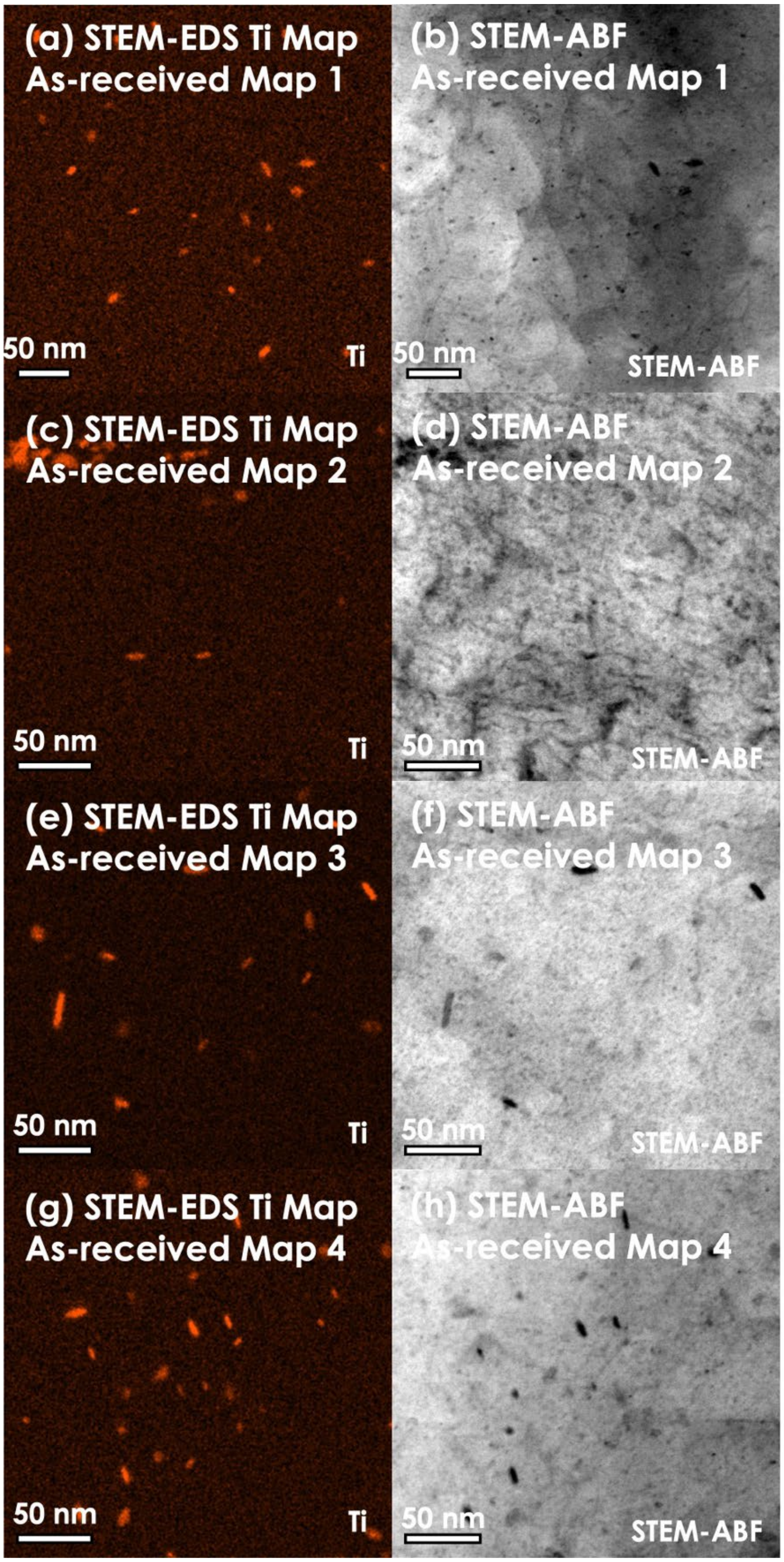

Figure S.2 STEM-EDS micrographs of Ti (a,c,e,g) and corresponding STEM-ABF micrographs (b,d,f,g) taken from the control CNA9 specimen, which was never irradiated or thermally annealed.

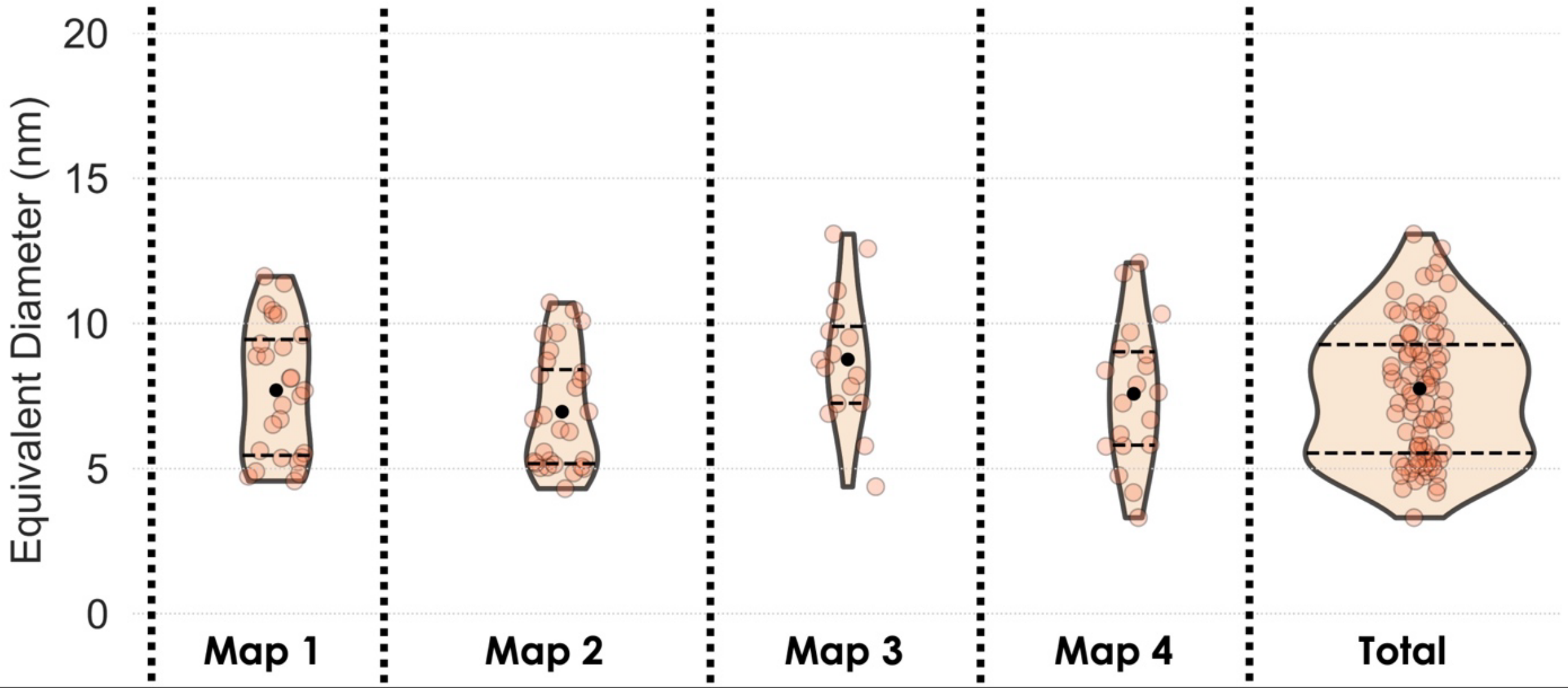

| | Map 1 | Map 2 | Map 3 | Map 4 | Total |
|---|---|---|---|---|---|
| N | 27 | 28 | 16 | 19 | **90** |
| $\rho$ ($m^{-3}$) | $1.6\times10^{21}$ | $3.1\times10^{21}$ | $2.0\times10^{21}$ | $2.4\times10^{21}$ | **$(2.3\pm0.3)\times10^{21}$** |
| $d_{eq}$(nm) | 7.7 | 7.0 | 8.8 | 7.6 | **7.8±0.3** |
| f | $4.9\times10^{-4}$ | $6.9\times10^{-4}$ | $8.3\times10^{-4}$ | $7.0\times10^{-4}$ | **$(6.8\pm0.6)\times10^{-4}$** |
| $\rho/\rho_{CTRL}$ | 0.7±0.09 | 0.7±0.09 | 1.4±0.2 | 0.9±0.1 | |
| $d/d_{CTRL}$ | 1.0±0.04 | 0.9±0.03 | 1.1±0.04 | 1.0±0.04 | |
| $f/f_{CTRL}$ | 0.7±0.07 | 0.7±0.06 | 1.0±0.09 | 1.2±0.1 | |

Figure S.3 Size distribution plots of precipitates counted in each STEM-EDS map of the control sample. Number of precipitates counted per map (N), number density of precipitates ($\rho$), average equivalent diameter ($d_{eq}$), and volume fraction (f) are shown for each map.

## Supplemental C: Precipitate Data

This supplemental section provides the summarized data and sample count for precipitates for each irradiation condition in Table S.1.

Table S.1 Data on the number of liftouts taken, the number of STEM-EDS maps taken, the number of precipitates counted (N), the number density of precipitates (ρ), the equivalent diameters of precipitates ($d_{eq}$), and the volume fraction of precipitates (f) for each irradiation condition.

| Sample | # of Liftouts Taken | # of EDS Maps Taken | N | ρ ($m^{-3}$) | $d_{eq}$(nm) | f |
|---|---|---|---|---|---|---|
| As-received | 1 | 5 | 90 | $(2.3\pm0.3)\times10^{21}$ | 7.8±0.3 | $(6.8\pm0.6)\times10^{-4}$ |
| 300°C, 15 dpa | 3 | 11 | 49 | $(0.9\pm0.2)\times10^{21}$ | 8.7±0.6 | $(3.9\pm1.0)\times10^{-4}$ |
| 300°C, 50 dpa | 1 | 5 | N.O. | N.O. | N.O. | N.O. |
| 400°C, 15 dpa | 3 | 9 | 67 | $(1.5\pm0.2)\times10^{21}$ | 7.6±0.4 | $(4.4\pm0.9)\times10^{-4}$ |
| 500°C, 1 dpa | 1 | 5 | 106 | $(3.3\pm0.9)\times10^{21}$ | 6.6±0.4 | $(6.9\pm0.2)\times10^{-4}$ |
| 500°C, 5 dpa | 1 | 5 | 86 | $(2.9\pm0.9)\times10^{21}$ | 7.4±0.4 | $(9.3\pm3.4)\times10^{-4}$ |
| 500°C, 15 dpa | 3 | 12 | 109 | $(1.7\pm0.3)\times10^{21}$ | 10.4±0.9 | $(11.1\pm2.3)\times10^{-4}$ |
| 500°C, 50 dpa | 2 | 9 | N.O. | N.O. | N.O. | N.O. |
| 500°C, 100 dpa | 2 | 6 | N.O. | N.O. | N.O. | N.O. |
| 600°C, 15 dpa | 3 | 9 | 69 | $(1.3\pm0.2)\times10^{21}$ | 14.9±2.2 | $(42.0\pm11.0)\times10^{-4}$ |

## Supplemental D: Ostwald ripening

Ostwald ripening is a coarsening process by which large particles grow at the expense of smaller ones [2, 3]. An analysis was performed to see if radiation-enhanced Ostwald ripening was occurring in the CNA9 samples irradiated at 500°C. As Ostwald ripening is a time-dependent phenomena, the 500°C series offered the best option for this analysis with three damage levels examined. The coarsening of particles under irradiation to $0.5_{\mathrm{TiC}}/1_{\mathrm{matrix}}$, $2.3_{\mathrm{TiC}}/5_{\mathrm{matrix}}$, $7_{\mathrm{TiC}}/15_{\mathrm{matrix}}$ dpa at 500°C was modeled using the equation for Ostwald ripening [2]:

$$r^3(t) - r_0^3(0) = \frac{8\sigma\Omega^2 D c_e}{9RT} \quad \text{Eq. S.1}$$

where $r_0^3(0)$ is the mean radius of the control specimen, $\sigma$ is the interfacial energy (J/m$^2$), Ω is the molar volume of the precipitate (m$^3$/mol), D is the diffusion coefficient of Ti (m$^2$/s), $C_e$ is the matrix concentration of Ti in equilibrium (atom fraction), $R$ is the universal gas constant J/mol-K), and $T$ is temperature (K). The thermal diffusion of the Ti solute atom occurs via the vacancy mechanism (Eq. S.2) and the diffusion of Ti under single beam irradiation can be calculated from the equation for radiation-enhanced diffusion (RED) of a solute (Eq. S.5).

$$D_{thermal}^{Ti} = D_{v,thermal}^{Ti} C_v^0 + D_{i,thermal}^{Ti} C_i^0 \quad \text{Eq. S.2}$$

$$D_{(v,i),thermal}^{Ti} = \alpha a^2 v exp\left(\frac{S_m^{v,i}}{k}\right) exp\left(\frac{-E_m^{v,i}}{kT}\right) \quad \text{Eq. S.3}$$

$$C_{v,i}^0 = exp\left(\frac{S_f^{v,\mathrm{i}}}{k}\right) exp\left(\frac{-E_f^{v,\mathrm{i}}}{kT}\right) \quad \text{Eq. S.4}$$

$$D_{RED}^{Ti} = D_{v,thermal}^{Ti} \frac{C_v^{irr}}{C_v^0} + D_{i,thermal}^{Ti} \frac{C_i^{irr}}{C_i^0} \quad \text{Eq. S.5}$$

Eq. S.1 can be used to calculate the mean precipitate radius as a function of time, $r^3(t)$. A plot of $r^3(t)$ versus $t$ can be seen in Figure S.4a. In this plot, time is represented by the damage level of the irradiations using a dose rate of 7×10$^{-4}$ dpa/s. A linear dependence of the $r^3(t)$ with $t$ is indicative that diffusion-controlled Ostwald ripening is occurring. In addition, a linear dependence of the $r^2(t)$ with $t$ is indicative that interface-controlled Ostwald ripening is occurring (Figure S.4b). In addition, a linear dependence with time would be expected for the inverse of the number density of precipitates if either kind of Ostwald ripening was occurring (Figure S.4c).

The simulated coarsening for the single beam irradiation calculated with $D_{RED}^{Ti}$ is shown with the dashed orange line in Figure S.4. The simulated thermal coarsening (*i.e.*, not under thermal ageing) calculated with $D_{thermal}^{Ti}$ is shown with the solid gray line in Figure S.4. The linear fits between the control specimen and the single beam $7_{TiC}/15_{matrix}$ dpa specimen for the radii and density changes are shown by the dashed blue curves. Figure S.4 shows that experimental values for the single beam set of irradiations to $0.5_{TiC}/1_{matrix}$, $2.3_{TiC}/5_{matrix}$, and $7_{TiC}/15_{matrix}$ dpa at 500°C (solid orange lines with markers) for the relationships of $r^3(t)$ versus t, $r^2(t)$ versus t, and $\rho^{-1}$ versus *t* deviate from linear behavior.

Thus, the deviation from linear behavior suggests the coarsening was not in line with Ostwald ripening or multiple complex mechanisms including Ostwald ripening are at play.

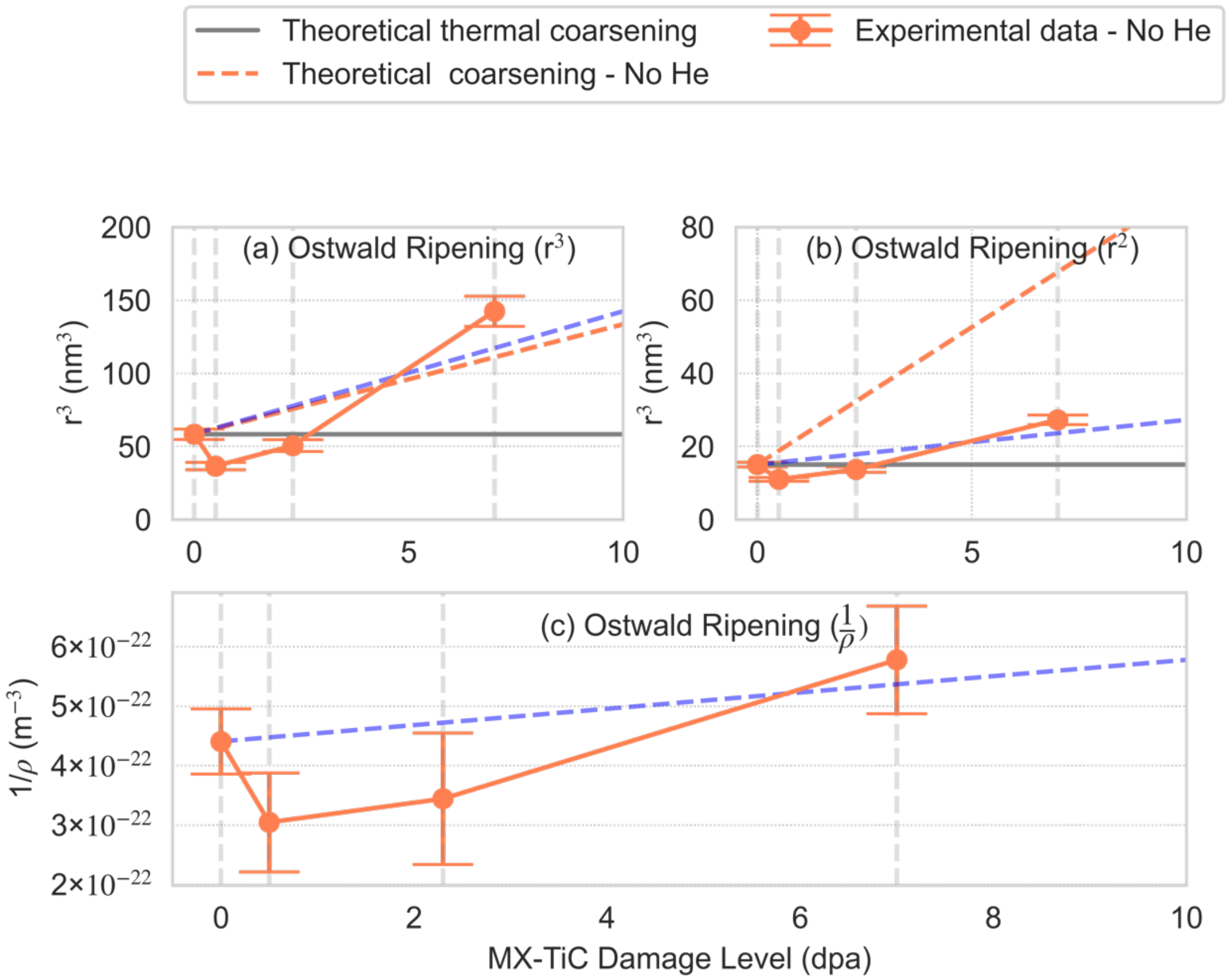

Figure S.4 Plots for (a) $r^3$ versus time (diffusion-controlled Ostwald ripening), (b) $r^2$ versus time (interface-controlled Ostwald ripening), and (c) $1/\rho$ versus t (generic Ostwald ripening) for the single beam conditions at 500°C.

**Supplemental E: 50 dpa thermal**

Figure S.5 shows the precipitates past the ion implantation depth in the 500°C condition to $23_{TiC}/50_{matrix}$ dpa. Precipitates were thermally stable during annealing for ~20 hours. Table S.2 shows the analysis for the precipitates in the thermally aged condition versus the control condition. There appears to be a significant decrease in the number density of precipitates with thermal ageing, but the equivalent diameter remained unchanged.

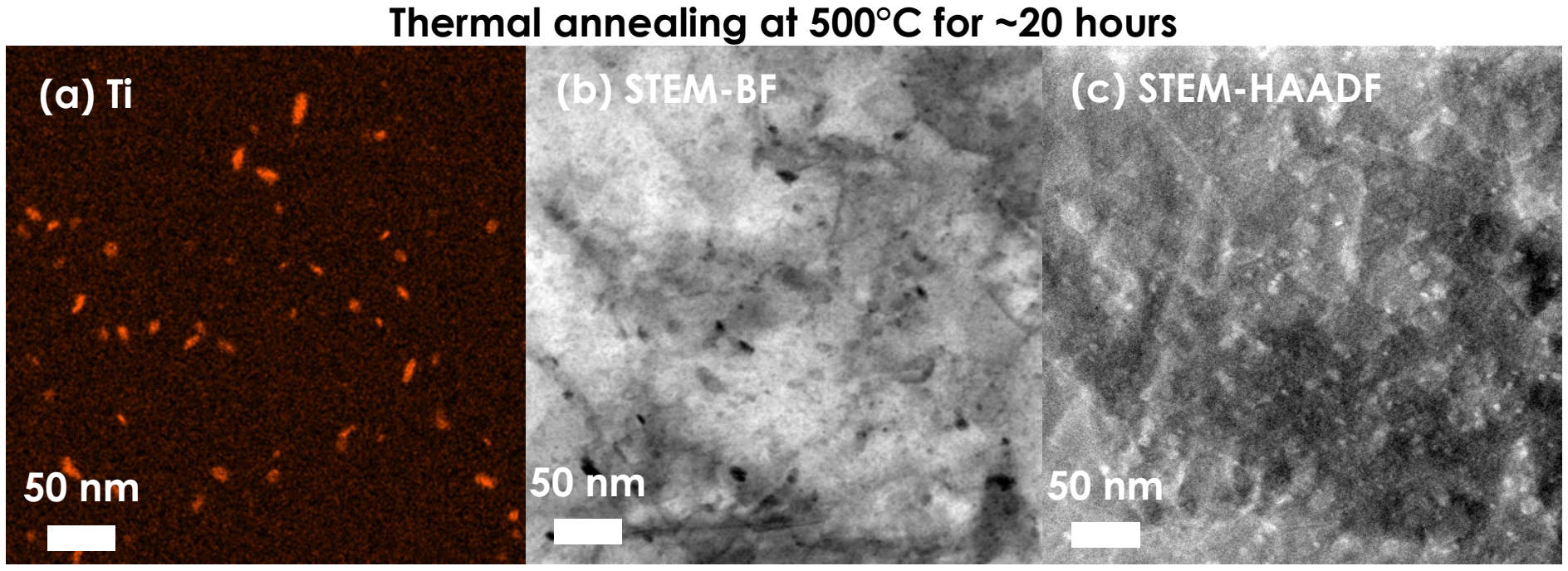

Figure S.5 (a) Example STEM-EDS map of Ti with corresponding (b) STEM-BF and (c) STEM-HAADF images of the sample thermally annealed for ~20 hours at 500°C.

Table S.2 The precipitate statistics for the thermally aged condition (500°C, ~20 hours) and the control condition.

| **Sample** | **N** | **ρ** | **d** | **f** |
|---|---|---|---|---|
| Control | 745 | (2.7±0.3) $10^{21}$ $m^{-3}$ | 7.9±0.3 nm | (9.0±1.4) $10^{-4}$ |
| Thermally aged 500°C ~20 hours | 74 | (1.3±0.4) $10^{21}$ $m^{-3}$ | 8.0±0.4 nm | (4.8±1.8) $10^{-4}$ |

**Supplemental F: Recoil Range Calculation of Ti Atoms from TiC Precipitates and Recoil Rate of TiC Precipitates using SRIM**

---

To estimate the recoil range of Ti atoms from TiC precipitates under irradiation, a SRIM analysis was performed. Multiple SRIM calculations were run to estimate recoil ranges of Ti atoms from various sized precipitates and precipitates located in different areas of the nominal damage region. The TiC precipitates were modeled as layers of Ti and C atoms. The nominal damage region occurred 1100-1300 nm beneath the surface of the bulk sample of CNA9, with the exact nominal damage and damage rate occurring at 1200 nm beneath the surface. The "Detailed Calculation with Full Damage Cascades" mode of SRIM with 1,000 ions was used in order to obtain the range of Ti atoms from the TiC layer. A displacement energy of 40 eV was used for Fe, Cr, Ni, V, and Mn, and 60 eV for Nb and Mo for both the initial simulation using Grade 91 and for the shorter CNA9 simulation. For all other elements, a displacement energy of 25 eV was used [4, 5].

First, a sensitivity analysis of the range dependence on TiC precipitate size was calculated. The SRIM calculation was set up as follows. A layer of CNA9 was simulated from the surface of the bulk sample to 1200 nm beneath the surface of the bulk sample. Then a layer of TiC was simulated, the layer having a width of 3, 5, 7, 10, or 90 nm. Another Layer of CNA9 was simulated from the end of the TiC layer to 2500 nm beneath the surface, where the 9 MeV $Fe^{3+}$ ions were shown to come to rest (Figure S.3). A TiC precipitate was modelled as a layer of 40% Ti and 60% C atoms. The ratio of Ti and C atoms in the layer was garnered from Thermo-Calc estimations of the ratio of Ti and C in the precipitates. The various widths simulated precipitates of various sizes. The values were chosen to represent the observed sizes for the MX-TiC and non-MX TiC precipitates in the control specimen.

The weighted average recoil range was then calculated from the SRIM results from each SRIM run. The weighted average of the recoil distance of a Ti solute outside of a TiC layer was calculated as follows, using a TiC layer located at 1200-1205 nm as an example:

$$R = \frac{\sum_{i=depth\ of\ 1205nm}^{depth\ at\ end\ of\ Ti\ range} counts\ in\ each\ bin/100\ \times\ (bin - 1205nm)}{\sum_{i=depth\ of\ 1205nm}^{depth\ at\ end\ of\ Ti\ range} counts\ in\ each\ bin/100} \qquad \text{Eq. S.6}$$

The counts represent the percentage of Ti in a bin. The bin represents the depth, and subtracting 1205 nm from the bin will obtain the projected range from the edge of the precipitate. The bin resolution was set by the viewing window parameter in SRIM and was equal to 0.3 nm. This way, the amount of Ti atom recoils within and outside the precipitate edge are calculated. As such, R represents the range outside the precipitate edge.

Table S.3 summarizes the results of this analysis. Figure S.6 shows the distributions of Ti atoms as a function of depth for TiC layers of 3, 5, 7, 10, and 90 nm. The weighted recoil ranges are approximately 0.5-0.7 nm. The recoil resolution efficiency was calculated as the ratio of the number of Ti atoms that recoiled outside the precipitate to those that recoiled within. As the MKI model suggests, this value increases for decreasing precipitate sizes. A layer of 7 nm was then chosen to continue with, as the ranges did not significantly change between the various layers and because the 7 nm represents the average size of the MX-TiC precipitates in the control specimen best, thus providing an average estimate of the range for MX-TiC precipitates.

Second, a sensitivity analysis of the range dependence on the location of the TiC layer was calculated. A 7 nm layer of TiC was placed at 1100-1107, 1200-1207, and 1293-1300 nm to cover the full range of the nominal damage region (Figure S.7). Table S.3 and Table S.4 display the results of these calculations. The weighted recoil ranges 0.9, 0.5, and 0.7 nm for layers at 1100-1107, 1200-1207, and 1293-1300 nm, respectively. The average of these values is 0.7 nm. This is the final value used for the range in the calculations.

Table S.3 Summary of results from SRIM calculations to obtain the recoil distance of Ti atoms from different sized precipitates under irradiation in CNA9.

| Width of TiC Layer | Location of Layer | Weighted Recoil Distance | Recoil Resolution Efficiency |
|---|---|---|---|
| 3 nm | 1200-1203 nm | 0.6 nm | 20.7% |
| 5 nm | 1200-1205 nm | 0.7 nm | 11.3% |
| 7 nm | 1200-1207 nm | 0.5 nm | 10.1% |
| 10 nm | 1200-1210 nm | 0.5 nm | 8.7% |
| 90 nm | 1200-1290 nm | 0.5 nm | 2.8% |

Table S.4 Summary of results from SRIM calculations to obtain the dependence of the recoil distance of Ti atoms from same sized precipitates located at different locations in the nominal damage region under irradiation in CNA9.

| Width of TiC Layer | Location of Layer | Weighted Recoil Distance | Recoil Resolution Efficiency |
|---|---|---|---|
| 7 nm | 1100-1107 nm | 0.9 nm | 15.2% |
| 7 nm | 1200-1207 nm | 0.5 nm | 10.1% |
| 7 nm | 1293-1300 nm | 0.7 nm | 17.5% |

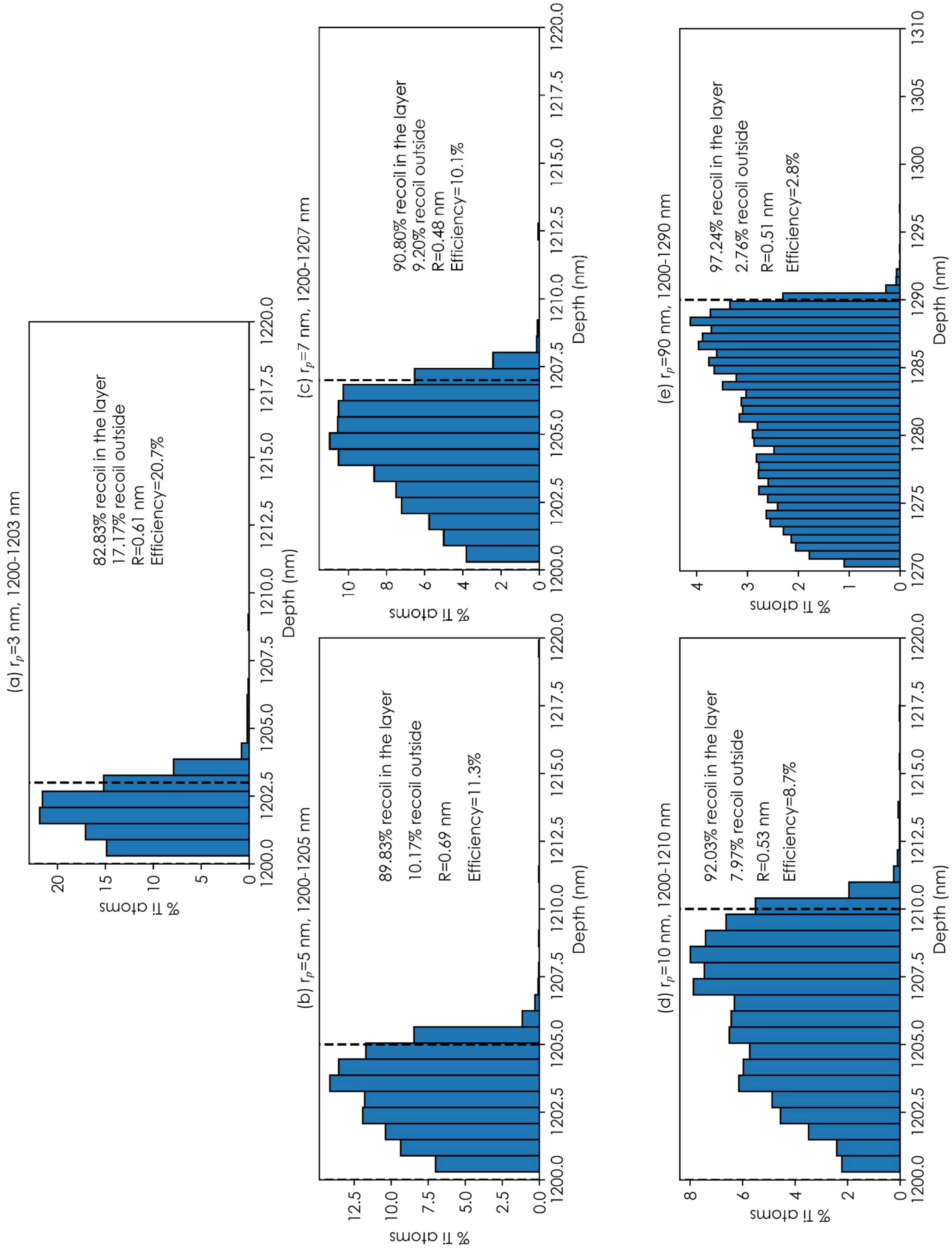

Figure S.6 SRIM analysis of recoiling Ti atoms from a TiC layer with widths of (a) 3nm, (b) 5nm, (c) 7 nm, (d) 10 nm, and (e) 90 nm.

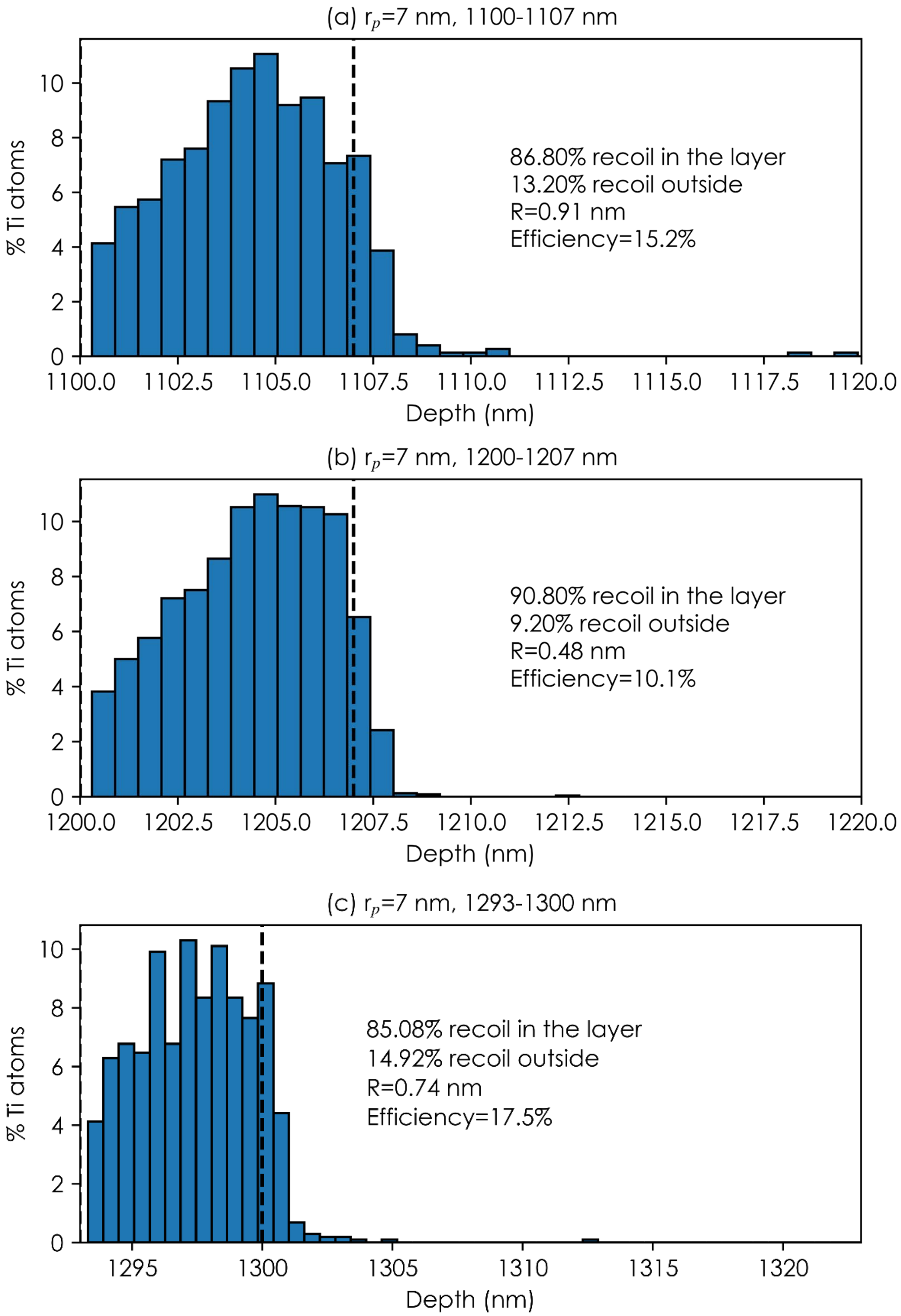

Figure S.7 SRIM analysis of 7 nm layers of TiC located at positions of (a) 1100-1107 nm, (b) 1200-1207 nm, and (c) 1293-1300 nm.

## Supplemental G: Determining coherency of MX-TiC Precipitates

Figure S. 8 shows the high-resolution (a) BF and (b) HAADF images with the corresponding (c) diffraction pattern of the small MX-TiC precipitates. It can be observed that the MX-TiC precipitates have a lattice parameter twice the size of the matrix and an FCC structure that had a cube-on-cube orientation relationship with the matrix. The broad faces of the precipitate are coherent with the matrix and the edges are incoherent or semi-coherent with the matrix. The d-spacings for $d_{\{220\}}^{TiC}$ and $d_{\{110\}}^{matrix}$ were identical but the d-spacing for $d_{\{200\}}^{TiC}$ was found to be twice as large as for that of the $d_{\{200\}}^{matrix}$. Hence, the parallel planes of the $\{200\}_{\text{matrix}}$ and the $\{200\}_{\text{TiC}}$ are the incoherent or semi-coherent interfaces of the TiC nanoprecipitates and the parallel planes of the $\{110\}_{\text{matrix}}$ and the $\{220\}_{\text{TiC}}$ are the coherent interfaces. For a plate-like precipitate like the MX-TiC precipitate, the coherent edges are assumed to induce large coherency strains, but there will not be such strains at the edges [6].

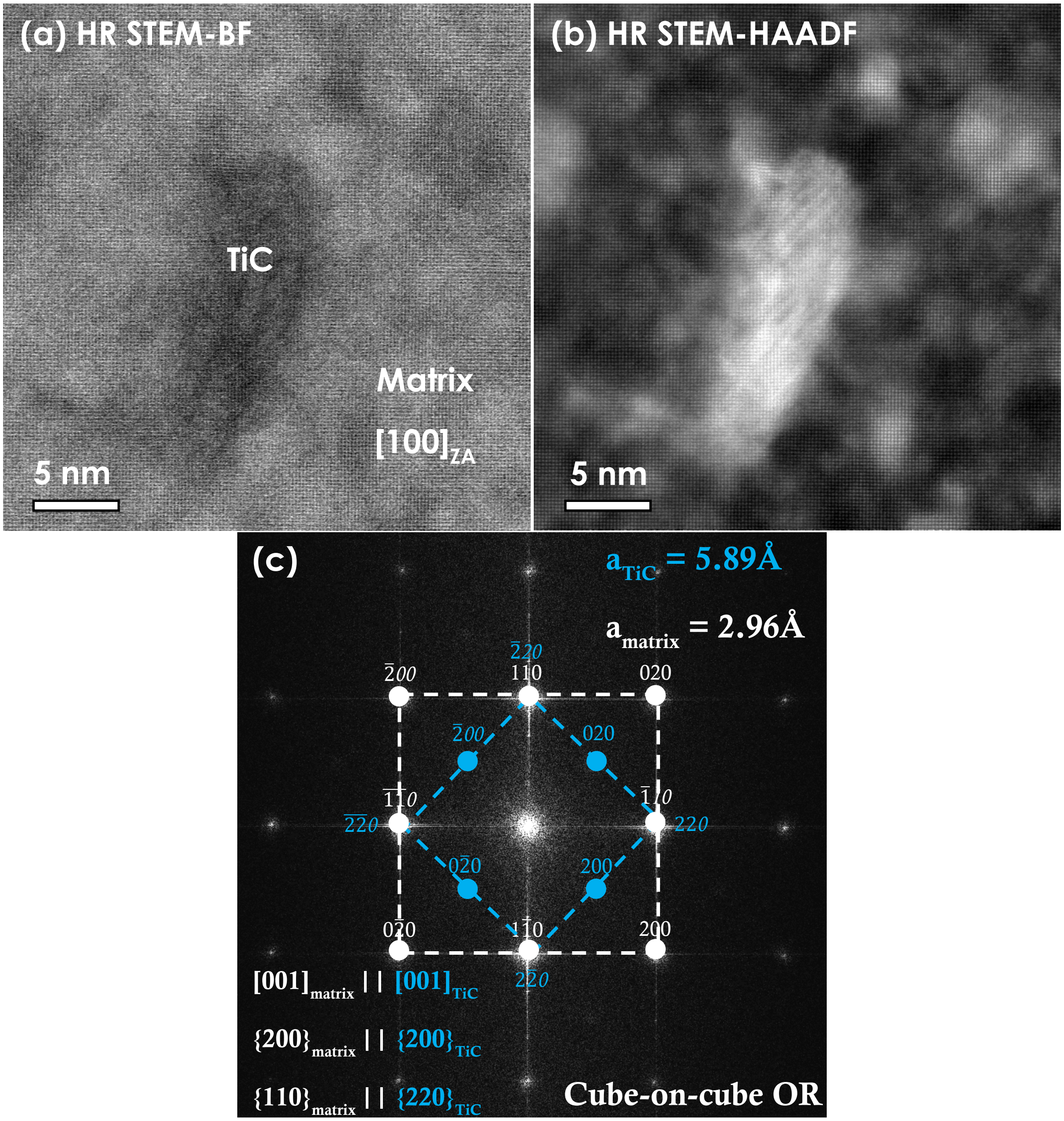

Figure S.8 (a) HR STEM-BF image, (b) STEM-HAADF image, and (c) the diffraction pattern of a MX-TiC nanoprecipitate from the control specimen of CNA9.

**Supplemental H: Determination of Sink Strength**

---

The total sink strengths ($k^2$) for interstitials and vacancies are the sum of the sink strengths of the individually measured sinks, accounting for the bias of interstitials for dislocations:

$$k_v^2 = \sum_s k_{vs}^2 = k_{gb}^2 + k_{ppts}^2 + k_{dis}^2 \qquad \text{Eq. S. 7}$$

$$k_i^2 = \sum_s k_{is}^2 = k_{gb}^2 + k_{ppts}^2 + k_{dis}^2 \qquad \text{Eq. S. 8}$$

$$k^2 = k_v^2 + k_i^2 \qquad \text{Eq. S. 9}$$

where $k_{gb}^2$ is the grain boundary sink strength from PAGs and laths, $k_{ppts}^2$ is the precipitate sink strength from the MX-TiC precipitates, and $k_{dis}^2$ is the dislocation sink strength.

The sink strength of MX-TiC precipitates was calculated as [7]:

$$k_{ppt}^2 = 4\pi\rho_{ppt}r_{ppt} \qquad \text{Eq. S. 10}$$

where $\rho_{ppt}$ is the number density of precipitates and $r_{ppt}$ is the average equivalent radius of the precipitates [4].

The sink strength of dislocations was calculated as:

$$k_{dis}^2 = k_{loop}^2 + k_{line}^2 \qquad \text{Eq. S. 11}$$

$$k_{(loop,line),i}^2 = 2B_i\pi\rho_{(loop,line)}r_{(loop,line)} \qquad \text{Eq. S.12}$$

$$k_{(loop,line),v}^2 = 2B_v\pi\rho_{(loop,line)}r_{(loop,line)} \qquad \text{Eq. S. 13}$$

$$k_{(loop,line)}^2 = k_{(loop,line),i}^2 + k_{(loop,line),v}^2 \qquad \text{Eq. S. 14}$$

where $k_{loop}^2$ is the sink strength of dislocation loops, $k_{line}^2$ is the sink strength of dislocation lines, $B_i$ is the bias factor of dislocations for interstitials, and $B_v$ is the bias factor of dislocations for vacancies. $B_v$ is assumed to be 1, meaning dislocations are not biased toward vacancies, based off of literature [4]. A value of 8% was used for $B_i$.

The sink strength of grain boundaries was calculated using the following equation:

$$k_{gb}^2 = k_{PAG}^2 + k_{lath}^2 \qquad \text{Eq. S. 15}$$

where

$$k^2_{PAG,lath} = \frac{6\sqrt{k^2_{inside\ grain}}}{d_{PAG,lath}} \quad \text{Eq. S. 16}$$

$$k^2_{inside\ grain} = k^2_{cav} + k^2_{ppts} + k^2_{dis}\ \text{(SI)} \quad \text{Eq. S. 17}$$

The measured values of precipitates from this work were used in Eq. S10. Dislocation loop size and density for 300°C was taken from Ref. [8]. Ref. [8] irradiated Grade 91 to 30 dpa at 300°C with a dose rate $2\times10^{-3}$ dpa/s. As dislocation loop size and density are a function of temperature, dose, and dose rate, it can be assumed that the value used here will not exactly match CNA9's value at 300°C but were appropriate for use due to lack of literature data [9]. Values for dislocation loop size and density and for dislocation line density for all temperatures besides 300°C were taken from Ref. [5]. Dislocation line density for 300°C assumed to be same as at 400°C, as data was lacking from literature. Values from Ref. [5] did not match the experimental conditions in this work exactly. Ref. [5] used Grade 91 irradiated at 406°C-16.6dpa-$7\times10^{-4}$ dpa/s-4.3 appm He/dpa, 480°C-16.6dap-$7\times10^{-4}$ dpa/s-4.3 appm He/dpa, and 570°C-15.4dpa-$7\times10^{-4}$ dpa/s-4.3 appm He/dpa. Hence, values from Ref. [5] input for 400°C in these calculations were taken from the experiment run at 406°C. Values input for 500°C in these calculations were taken from the experiment run at 480°C. Values input for 600°C in these calculations were taken from the experiment run at 570°C. Errors for dislocation loop size and density from Ref. [5] were reported in the reference. Grade 91 is an appropriate surrogate material due to the similarity in composition to CNA9 and the similar grain and lattice structures. Hence, it is assumed that values of dislocation line densities of Grade 91 can be used for CNA9, within appropriate reason. Values for the prior austenite grain (PAG) size was taken from Ref. [5] and the martensite lath size was taken from Ref. [10]. Error for the lath size was assumed to be 10% to cover the spread of lath sizes found in literature [11].

Table S.5 Values input to calculate sink strength for single ion irradiation conditions in the temperature series to 15 dpa with 7×10$^{-4}$ dpa/s. N.C. means not calculated.

| | | | | |
|---|---|---|---|---|
| Dislocation Lines | Number density (m$^{-3}$) | 300°C: 3.8×10$^{14}$<br>400°C: 3.8×10$^{14}$<br>500°C: 0.4×10$^{14}$<br>600°C: N.O. | $k_{line}^2$ (m$^{-2}$) | 300°C: 7.9×10$^{14}$<br>400°C: 7.9×10$^{14}$<br>500°C: 8.7×10$^{13}$<br>600°C: N.C. |
| Dislocation Loops | Number density (m$^{-3}$) | 300°C: 4.1×10$^{22}$<br>400°C: 12×10$^{21}$<br>500°C: 0.46×10$^{22}$<br>600°C: N.O. | $k_{loop}^2$ (m$^{-2}$) | 300°C: (2.3±1.1)×10$^{15}$<br>400°C: (1.6±0.2)×10$^{15}$<br>500°C: (1.7±0.2)×10$^{14}$<br>600°C: N.C. |
| | Radius (nm) | 300°C: 4.3±2.1<br>400°C: 10.1±1.2<br>500°C: 29±3.8<br>600°C: N.O. | | |
| MX-TiC precipitates | Number density (m$^{-3}$) | 300°C: (8.8±1.8)×10$^{20}$<br>400°C: (1.4±0.2)×10$^{21}$<br>500°C: (1.7±0.3)×10$^{21}$<br>600°C: (1.3±0.2)×10$^{21}$ | $k_{MX}^2$ (m$^{-2}$) | 300°C: (4.8±1.0)×10$^{13}$<br>400°C: (6.9±1.0)×10$^{13}$<br>500°C: (1.1±0.2)×10$^{14}$<br>600°C: (1.2±0.2)×10$^{14}$ |
| | Radius (nm) | 300°C: 4.4±0.09<br>400°C: 3.8±0.06<br>500°C: 5.2±0.1<br>600°C: 7.4±0.4 | | |
| Grain Boundaries | Diameter of PAGs (µm) | 15 | $k_{GB}^2$ (m$^{-2}$) | 300°C: (7.0±0.7)×10$^{14}$<br>400°C: (6.1±0.6)×10$^{14}$<br>500°C: (2.4±0.2)×10$^{14}$<br>600°C: (1.4±0.1)×10$^{14}$ |
| | Diameter of laths (nm) | 500±50 | | |
| Total | | | $k_{tot}^2$ (m$^{-2}$) | 300°C: (4.6±1.1)×10$^{15}$<br>400°C: (3.7±0.2)×10$^{15}$<br>500°C: (9.7±0.6)×10$^{14}$<br>600°C: (5.1±0.5)×10$^{14}$ |